\documentclass[12pt,twoside]{article}
\usepackage{latexsym}
\usepackage{longtable}
\usepackage{epsfig}
\usepackage{graphicx,psfrag}
\usepackage{amssymb,amsmath,amsthm,booktabs,mathtools}
\usepackage{bm}
\usepackage{color}
\usepackage{subfig}
\usepackage{dutchcal}
\usepackage{yfonts}

\newtheorem{rema}{Remark}[section]

\DeclareMathOperator{\Tr}{Tr}

\newcommand{\bc}{\begin{center}}
\newcommand{\ec}{\end{center}}
\def\ba#1{\begin{array}{#1}\displaystyle}
\newcommand{\ea}{\end{array}}

\newcommand{\beq}{\begin{equation}}
\newcommand{\eeq}{\end{equation}}
\newcommand{\beqa}{\begin{eqnarray}}
\newcommand{\eeqa}{\end{eqnarray}}
\newcommand{\no}{\nonumber}
\newcommand{\n}{\nonumber\\}
\newcommand{\bi}{\begin{itemize}}
\newcommand{\ei}{\end{itemize}}

\def\lt#1{\left#1}
\def\rt#1{\right#1}
\def\t#1{\tilde{#1}}

\def\b#1{\bar{#1}}
\def\frc#1#2{\frac{#1}{#2}}

\newcommand{\p}{\partial}

\newcommand{\bra}{\langle}
\newcommand{\ket}{\rangle}
\newcommand{\Z}{{\mathbb{Z}}}
\newcommand{\N}{{\mathbb{N}}}
\newcommand{\R}{{\mathbb{R}}}

\newcommand{\ep}{\epsilon}
\newcommand{\varep}{\varepsilon}

\newcommand{\re}{{\rm e}}

\newcommand{\dd}{{\rm d}}
\DeclareMathOperator{\sgn}{sgn}

\def\gas#1{\mathcal{#1}}
\def\chain#1{\mathcal{#1}^{\rm c}}
\def\average#1{\mathsf{#1}}
\def\averagec#1{\mathsf{#1}^{\rm c}}
\def\magas#1{\mathsf{#1}}
\def\machain#1{\mathsf{#1}^{\rm c}}

\newcommand{\halmos}{\rule{1ex}{1.4ex}}
\newcommand{\eproof}{\hspace*{\fill}\mbox{$\halmos$}}

\begin{document}

\begin{center}
{\Large {\bf Generalised hydrodynamics of the  \\[0.2cm] classical Toda system}}

\vspace{1cm}

{\large Benjamin Doyon}
\vspace{0.2cm}

{\small\em
Department of Mathematics, King's College London, Strand, London WC2R 2LS, U.K.}
\end{center}
\vspace{1cm}

\noindent We obtain the exact generalised hydrodynamics for the integrable Toda system. The Toda system can be seen in a dual way, both as a gas and as a chain. In the gas point of view, using the elastic and factorised scattering of Toda particles, we obtain the generalised free energy and exact average currents, and write down the Euler hydrodynamic equations. This is written both as a continuity equation for the density of asymptotic momenta, and in terms of normal modes. This is based on the classical thermodynamic Bethe ansatz (TBA), with a single quasiparticle type of Boltzmann statistics. By explicitly connecting chain and gas conserved densities and currents, we then derive the thermodynamics and hydrodynamics of the chain. As the gas and chain have different notions of length, they have different hydrodynamics, and in particular the velocities of normal modes differ. We also give a derivation of the classical TBA equations for the gas thermodynamics from the factorised scattering of Toda particles. 
\vspace{0.5cm}

{\ }\hfill
\today

\section{Introduction}

Recently there has been a lot of activity in the study of many-body systems out of equilibrium. In one dimension,  integrability has a striking effect on non-equilibrium physics, because the presence of infinitely many conserved quantities and ballistic currents prevent the usual thermalisation processes to occur. In the quantum context, works on quantum quenches have led to the notion of generalised Gibbs ensembles (GGEs) \cite{PhysRevLett.98.050405} (see the reviews \cite{eisertreview,1742-5468-2016-6-064002}). It is found that large, isolated, homogeneous integrable systems locally relax to states that maximise entropy with respect to the infinity of available local and quasi-local \cite{PhysRevLett.115.157201,IlievskietalQuasilocal} conserved quantities. For inhomogeneous and non-stationary states, one uses local entropy maximisation as the core principle of hydrodynamics \cite{Spohn-book}.  Combined with GGEs, this gives generalised hydrodynamics (GHD) \cite{PhysRevX.6.041065,PhysRevLett.117.207201,DY,dNBD}.

GGEs and GHD are most conveniently formulated in the language of the thermodynamic Bethe ansatz (TBA) \cite{Yang-Yang-1969,ZAMOLODCHIKOV1990695}. Despite originally for quantum models, TBA is a flexible language based on scattering theory, adaptable to classical models, see \cite{T84,O85,TI85,TI86,CJF86, TSPCB86, BPT86,TBP86,BPT86} for equilibrium thermodynamics, \cite{Zakharov,El-2003,El-Kamchatnov-2005,El2011,CDE16} for hydrodynamics of soliton gases, and \cite{dLM16,Doyon-Spohn-HR-DomainWall,DYC18,10.21468/SciPostPhys.4.6.045} for the modern notions of classical GGEs and GHD. A powerful and general framework for the thermodynamics and hydrodynamics of integrable models is therefore proposed (see the discussions in \cite{10.21468/SciPostPhys.4.6.045,doyoncorrelations,dNBD2}).

In this paper, we construct the GGEs and Euler GHD for the Toda system \cite{TodaBook}. The goal is to illustrate GHD in a classical interacting particle model that is simple enough yet non-trivial, and that is in principle amenable to accurate numerical simulations \cite{KD16}. The main new results are the exact GGE averages of conserved currents, the Euler hydrodynamic equations, and the fluid normal modes. By considering appropriate scattering states, this paper also serves to emphasise that the structure of the TBA depends on the choice of vacuum, and that a good choice may reduce the complexity of the calculations. We finally show how the classical TBA can be derived from first principles.

The Toda system can be seen either as a chain of  ``displacements" with nearest-neighbour interaction as originally conceived by Toda \cite{To67}, or as a gas (or fluid -- no low-density assumption is made) of labelled Galilean particles moving in $\R$ with exponential repulsion \cite{Mo75}. The two viewpoints lead to slightly different hydrodynamics: the notion of length in the chain -- the number of sites -- is different from that in the gas -- the distance between the particles' positions. There is a duality whereby total ``charge" and ``volume" are exchanged: the number of gas particles is the chain volume, while the sum of displacements, a chain topological charge, is the gas volume. Accordingly, we consider two thermodynamic ensembles dual to each other: in the gas language, these are fixed chemical potential and volume (ensemble ${\cal E}_{\rm Landau}$), and fixed pressure and number of particles (ensemble ${\cal E}_{\rm Gibbs}$). We construct GGEs in both languages of ${\cal E}_{\rm Landau}$ and ${\cal E}_{\rm Gibbs}$, and we develop the GHD for both the gas and the chain.

The techniques we use are based on a classical scattering analysis. The scattering states above the chain's vacuum are formed by soliton waves and radiative corrections \cite{ZK65,KT}. We consider instead the simpler scattering states above the gas's vacuum, which are the asymptotic, freely propagating Toda particles themselves \cite{Mo75}. The scattering of Toda particles is elastic and factorises, and the exact scattering shift can be evaluated by elementary calculations. The asymptotic momenta are the eigenvalues of the Toda Lax matrix \cite{Fla74}. Inserting this in the Boltzmann-statistics, Galilean TBA gives the GGE specific Landau free energy. Following \cite{PhysRevX.6.041065,PhysRevLett.117.207201,10.21468/SciPostPhys.4.6.045}, the exact gas currents are also written, giving the GHD for the gas. The picture is similar to that of the hard rod \cite{Boldrighini1983,Spohn-book} or classical soliton gases or ``flea gases" \cite{Zakharov,CDE16,DYC18}: an effective velocity emerges due to the scattering shift of the Toda particles.

Using the equivalence of ensembles, we then establish the relation between the averages of currents in the gas and the chain. From this, we develop the chain GGEs and GHD. There too, there is a natural interpretation of the effective velocity using particle collisions.

Finally, we provide a first-principle derivation of the TBA free energy formula from the definition of the ensemble ${\cal E}_{\rm Landau}$ and the classical scattering of the Toda particles. We recast the partition function into multiple integrals over momenta of asymptotic particles. We evaluate the entropy (``state density") associated to a given set of momenta in terms of the scattering shift, and deduce the classical TBA equations from a mean-field approximation. The derivation is not specific to the Toda gas, and, with slight extensions, should apply to classical integrable systems more generally.

The classical TBA of various models has been derived from a semi-classical analysis of the quantum TBA, see the review \cite{dLM16}, and from the inverse scattering method, see e.g. \cite{TI85,TSPCB86, BPT86}. The derivation we present is related to, but  different from, the latter. The thermodynamics of the classical Toda system has been studied, see the review \cite{CLSTV94}.  Theodorakopoulos \cite{T84} and Opper \cite{O85} took a semi-classical limit of the quantum Toda TBA, obtaining the equilibrium thermodynamics of the classical model. It is a simple matter to extend their result to GGEs. The gas GGE we have obtained agrees with this semi-classical result. A more in-depth semi-classical analysis has been provided in \cite{BCM19}, which appeared shortly after the first version of the present paper.

Interestingly, H. Spohn has observed \cite{SpoToda} that the ensemble ${\cal E}_{\rm Gibbs}$ for the Toda chain is related to the Dumitriu-Edelman {\em random matrix theory} model \cite{DE02}. By taking an appropriate limit of the Dumitriu-Edelman model, the specific Gibbs free energy can be obtained. This gives a direct relation between the thermodynamics of integrable systems and random matrix theory, and an almost rigorous derivation of the Toda thermodynamics itself. The chain GGE we have obtained agrees with this random matrix theory approach, and the Dumitriu-Edelman eigenvalues are related to the asymptotic Toda momenta.

The paper is organised as follows. In section \ref{sectrel}, we discuss the model and establish the relation between the gas and the chain. In section \ref{sectghd}, we discuss the scattering of Toda particles and develop the thermodynamics and hydrodynamics. In section \ref{secttba}, we provide our derivation of the classical TBA from Toda scattering. We conclude in section \ref{sectconclu}.

\section{Relation between the Toda gas and chain} \label{sectrel}

\subsection{Hamiltonian and thermodynamic limit}\label{ssecthamil}

Let $N\in\N=\{1,2,3,\ldots\}$. For every $m\in \Z$ let $p(m)$ and $x(m)$ be independent canonical variables, with Poisson bracket $\{p(m),x(m')\} = \delta_{m,m'}$. For $m>N$ and $m<1$, set $p(m)=0$ and $x(m) = (m-1)\Lambda$ with $\Lambda\to\infty$ -- these are non-dynamical and used here for notational convenience (see \cite{Mo75}). Let the hamiltonian be
\beq\label{H}
	H = \sum_{m\in \Z\cap[1,N]} \lt(\frc12 p(m)^2+ \re^{-r(m)}\rt)
	= \sum_{m\in\Z} \lt(\frc12 p(m)^2+ \re^{-r(m)}\rt),\quad
	r(m) = x(m+1)-x(m).
\eeq

This model can be seen in two different ways. It is first the open {\em classical Toda chain}, with $N$ sites $m\in \Z\cap[1,N]$ and on each site the conjugate dynamical variables $p(m)\in\R$ and $x(m)\in\R$. The choice of open boundary conditions is for convenience of the discussion; as we will be considering the thermodynamic limit, the results are expected to hold as well, in this limit, for the periodic chain. It is also the {\em classical Toda gas}, with $N$ particles parametrised by $m$ lying at positions $x(m)$ and having conjugate momenta $p(m)$. In the gas viewpoint, only particles with neighbouring labels interact, and they repel each other in the direction that tends to keep the particle ordering on the line the same as the label ordering. In the chain viewpoint, only neighbouring sites interact, and the chain tends to ``slant", separating the dynamical variables $x(m)$ as much as possible with a positive slope. In particular, there is no solution that reaches the infimum energy (which is 0) of the hamiltonian $H$.

In the viewpoint of a gas, the number of particles is
\beq
	N = \sum_{m\in \Z\cap[1,N]} 1
\eeq
and we may define the ``oriented volume" as
\beq\label{volume}
	R = x(N) - x(1) = \sum_{m\in  \Z\cap[1,N-1]} r(m).
\eeq
This variable can take positive and negative values, see below.

In the chain viewpoint, by contrast, $N$ is the volume, and $R$ is a {\em topological charge}, depending on the end-point dynamical variables only. In this viewpoint, one usually includes a pressure term in the hamiltonian itself, by the addition of $Pr(m)$ for some $P>0$ within the sum in \eqref{H}. With this term, there is a lowest-energy solution. The topological charge $R$ is related to a ``counting" of Toda chain's solitons, much like $N$ counts particles in the gas viewpoint.  For instance, at $P=1$, in the lowest energy solution, all particles stand at the same position, and in this case, $R$ is the total strength of solitons in multi-soliton solutions of the Toda chain. However, to be precise, in a generic solution one must also account for radiative corrections, which contribute to $R$. Besides making some comments, we will not discuss further solitons and radiative corrections.

We will be interested in taking the thermodynamics limit, constructing states with, almost surely,
\beq\label{thermo}
	N,|R|\to\infty,\qquad R/N = \nu\ \mbox{finite and fixed.}
\eeq
The states are thermodynamic ensembles, where the pressure $P>0$ is seen as the thermodynamic dual to the oriented volume $R$, and is sufficient to confine the gas. With strong enough pressure, the variable $\nu$ may take negative values. The quantity $|R|$ from \eqref{volume} is in any case a good approximation of the actual volume occupied by the gas in all states of interest; it gives the exact scaling of the actual volume with $N$ in the thermodynamic limit. In the gas picture, $|\nu|^{-1}$ is the particle density per unit length and $\sgn(\nu)$ is the approximate ordering of the particles' positions with respect to their labels in the state of interest; we will refer to $\sgn(\nu)$ as the ordering signature. In the chain picture, $\nu$ is the topological charge per site. In order to have a more uniform description, in discussing average densities in the gas picture, it will be convenient to consider ``oriented lengths", with for instance the quantity $\nu^{-1}$ being the average particle density per unit oriented length. Note that at $\nu=0$, there is a seeming singularity in the gas thermodynamics, as all particles lie within a ``small" volume; however this can be removed by appropriate choice of coordinates. There is no singularity of the chain thermodynamics at $\nu=0$.

The model has a natural symmetry under simultaneous shift of all $x(m)$, translation symmetry in the gas picture. This is a gauge degree of freedom, as it forms a non-compact group, and we fix it in order to define the state and to make sure that, in the thermodynamic limit, finite labels correspond to finite particles' positions. Explicitly, for convenience, we may fix this gauge by, on each configuration, first setting $x(1) = -\nu N/2$ and then {\em re-labelling} the dynamical variables $m\mapsto m+\Delta$, $\Delta\in\Z$ (translation in the chain), in such a way that $x(0)\geq 0,\ \nexists\, m\in\Z: 0<x(m)<x(0)$; that is, 0 labels the first particle just to the right of the position 0. The particles' positions are almost surely non-coincident, hence this defines uniquely the shift $\Delta$. Then, in the states of interest, we expect that almost surely, after the thermodynamic limit, for every $m\in\Z$ we have $x(m)\in \R$ and $p(m)\in\R$: positions and momenta are finite for finite labels.

Further, in the arguments below, it is convenient to refer to certain aspects of the distribution of positions which we expect to hold in the states of interest. First, we expect the random variables $x(m+1)-x(m)$ to have a distribution which is peaked strongly enough around $\nu$ -- for instance with exponential decay. In particular, it is very unlikely that the $x(m)$'s not be ordered on large scales. Second, we expect that there exist $\alpha\in[0,1)$ such that the scaled variables $(x(m)-\nu m)/m^\alpha$ have a limit in distribution at large $|m|$, which is also peaked strongly enough around 0. That is, at large label, $x(m)$ stays around the position $\nu m$ with fluctuations that are subleading. If the random variables $x(m+1)-x(m)$ were i.i.d., with $\nu$ the average, then $(x(m)-\nu m)/\sqrt{m}$ would tend to a Gaussian in distribution at large $m$
and so $\alpha=1/2$ is the natural guess (by the law of large numbers) -- but this precise value is not necessary for the discussion.

The equations of motion from the classical hamiltonian \eqref{H} are
\beq
	\frc{\dd x(m,t)}{\dd t}  = p(m,t),\quad \frc{\dd p(m,t)}{\dd t} = \re^{-(x(m,t)-x(m-1,t))} - \re^{-(x(m+1,t)-x(m,t))}.
\eeq
The thermodynamic states we will consider are stationary, and it is clear, from these equations of motion, that the properties discussed above are preserved in time.

\subsection{Conserved quantities}

Before giving a thermodynamic definition of the states, we need to construct the conserved quantities of the model. 
The Toda system is integrable, and possesses an infinite number of conserved quantities. A clear way of expressing these conserved quantities is using the Toda system's Lax pair found in \cite{Fla74}. The Lax matrix can be thought of as the infinite-dimensional matrix
\beq
	L = \lt(\begin{matrix}
	\ddots & \vdots  &  &  &  \vdots &  \\
	 \cdots & b(-1) & a(-1) & 0 & 0 & \cdots\\
	 & a(-1) & b(0) & a(0) & 0 &  \\
	 & 0 & a(0) & b(1) & a(1) &  \\
	 \cdots & 0 & 0 & a(1) & b(2) &  \cdots \\
	 & \vdots & & & \vdots &\ddots 
	\end{matrix}\rt)
\eeq
where $a(m) = \re^{-r(m)/2}$ and $b(m) = p(m)$ (we use a different normalisation than \cite{Fla74}). Because of the prescription on $p(m)$ and $x(m)$ for $m>N$ and $m<1$, the matrix is asymptotically zero on both directions along the diagonal. It satisfies
\beq\label{lax}
	\frc{\dd L}{\dd t} = [M,L]
\eeq
for appropriate $M$. As a consequence the trace of its powers 
\beq
	Q_i = 2^{1-i} \Tr (L^i),\quad i=1,2,3,\ldots.
\eeq
are conserved,
\beq\label{conscharge}
	\frc{\dd Q_i}{\dd t} = 0.
\eeq

As is clear by inspection, these are local conserved charges: that is, for each $i$, the quantity $Q_i$ is expressed as a sum over chain sites of local chain observables. More precisely, we say that a chain observable $\chain o(m,t)$ is local if there exists $\ell>0$ such that, for all $m$, $\chain o(m,t)$ is a function only of dynamical variables $x(m',t)$, $p(m',t)$ with labels $m':|m-m'|<\ell$. Then we can write
\beq\label{chargechain}
	Q_i = \sum_{m\in \Z} \chain{q}_i(m,t)
\eeq
where $\chain{q}_i(m,t)$ is a local chain observable at $m$. For instance,
\beq
	\chain{q}_1(m,t) = p(m,t),\qquad \chain{q}_2(m,t) = \frc{p(m,t)^2}2 + \re^{-r(m,t)}
\eeq
are the momentum and energy per site, respectively: $Q_1$ is the total momentum, and $Q_2=H$ is the total energy. These expressions are valid for finite $N$ under the prescription above Eq. \eqref{H}. Taking the thermodynamic limit, generically, the $Q_i$'s are almost surely infinite. However, they have finite densities. 

By locality and conservation, there exist local currents $\chain{j}_i(m)$ which satisfy
\beq\label{conschain}
	\frc{\dd}{\dd t}\chain{q}_i(m,t) + \chain{j}_i(m+1,t)-\chain{j}_i(m,t) = 0 \quad \forall\ m,i.
\eeq
These can be written explicitly, see \cite{SpoToda}.

It is also possible to write the conserved quantities $Q_i$ in terms of the local gas observables instead of local chain observables. Let $\gas o(x,t)$ be an observable of the Toda gas. Suppose there exists a local chain observable $\chain o(m,t)$ and a random almost surely finite set $M(x,t)\subset \Z$, such that the variable $\varep(x,t) = {\rm max}(|x- x(m,t)|:m\in M(x,t)\}$ has probability distribution that decays fast enough at large values, and $\gas o(x,t)$ can be written as
\beq\label{localgas}
	\gas o(x,t) = \sum_{m\in M(x,t)} \chain o(m,t).
\eeq
Then $\gas o(x,t)$ is a local gas observable. In general, a local gas observable is a finite sum of such expressions. We then define the local gas densities as
\beq\label{densgas}
	\gas q_i(x,t) = \sum_{m\in \Z} \chain q_i(m,t) \delta(x-x(m,t)).
\eeq
This is a density per unit length. We may replace Dirac's delta function by any finitely-supported function that integrates to one, for instance
\beq\label{repl}
	\delta(x-x(m,t)) \mapsto \frc{\Theta(\varep-|x- x(m,t)|)}{2\varep }
\eeq
for any fixed $\varep>0$, and we have the form \eqref{localgas} with $M = \{m:|x-x(m,t)|<\varep\}$. These local gas densities satisfy
\beq\label{chargegas}
	Q_i = \int_\R  \dd x\,\gas q_i(x,t).
\eeq

Again, by conservation and locality there are local currents,
\beq\label{consgas}
	\frc{\p}{\p t} \gas q_i(x,t) + \frc{\p}{\p x} \gas j_i(x,t) =0.
\eeq
A straightforward calculation shows that the currents take the form
\beq\label{curgas}
	\gas j_i(x,t) = \sum_{m}
	\big[\chain q_i(m,t)p(m,t)\delta(x-x(m,t)) + \chain j_i(m,t)(\Theta(x-x(m-1,t)) - \Theta(x-x(m,t)))\big].
\eeq
Integrated over smooth functions of $x$ with compact support, these are well-defined observables which only take finite values \cite{LLL77}. The first term in the summand on the right-hand side is clearly a local gas observable. The second term is also local. An argument is as follows. This term can be written as $\chain j_i(m,t)(\delta_{m\in S_+(x,t)}-\delta_{m\in S_-(x,t)})$ with $S_+(x,t)=\{m:x(m-1,t)<x,x(m,t)>x\}$ and $S_-(x,t)= \{m:x(m-1,t)>x,x(m,t)<x\}$. Define $\varep(x,t) = {\rm max}(|x(m,t)-x|:m\in S_+(x,t)\cup S_-(x,t))$. Because each $x(m+1,t)-x(m,t)$ is peaked strongly enough around $\nu$, the variable $\varep$ has a probability distribution that decays fast enough at large values.  Therefore we have the form \eqref{localgas}.

We also introduce a convenient notation for the number of particles and the oriented volume, along with their associated conserved densities:
\beq\label{def0}
	Q_0 = N,\quad \chain{q}_0(m,t) = 1\qquad\mbox{and}\qquad
	Q_{\b 0} = R,\quad \chain{q}_{\b 0}(m,t) = r(m,t)
\eeq
where the chain-local densities are expressed in the bulk, boundary effects being ignored.  It is a simple matter to verify that the associated currents are related to the momentum densities as
\beq\label{cur0}
	\gas{j}_0(x,t) = \gas{q}_1(x,t),\qquad
	\chain{j}_{\b 0}(m,t) = -\chain{q}_1(m,t),
\eeq
while $\chain{j}_0=0$ and $\gas{j}_{\b 0}$ has a more complicated expression. The index set for the conserved quantities is therefore extended to $\{0,\b 0,1,2,3,\ldots\}$.

We note that the first equation in \eqref{cur0} is just the expression of Galilean invariance in the gas picture, which states that the mass current equals the momentum density (here the mass is unity). The second equation can be interpreted as a similar statement, but for ``particles" being the total topological charge of the Toda chain. The sign is intuitively natural in the soliton picture: a negative amount is added to $Q_{\b 0}$ for a soliton joining large values of $x(m)-\nu m$ to small values of $x(m')-\nu m'$, $m\ll m'$, and since positive momenta $p(m'')$, $m<m''<m'$ correspond to a displacement of this soliton towards the right, the ``soliton momentum" at chain site $m$ is the negative of the $m$th particle's momentum.

\subsection{Generalised Gibbs ensembles and thermodynamic potentials}\label{ssectgge}

We are interested in the generalised Gibbs ensembles (GGEs), where the thermodynamic ensemble involves not only the hamiltonian $H$, but potentially all conserved quantities $Q_i$. We may consider the power series expansion $W=\sum_{i=1}^\infty \beta_i Q_i = \sum_{i=1}^{\infty} 2^{1-i}\beta_i \Tr(L^i) $, or more generally, following the ideas appearing in the thermodynamic Bethe ansatz description of GGEs \cite{1742-5468-2016-6-064002}, the trace of an arbitrary function of $L$, say $w(L)$, which may or may not have a Taylor series expansion $w(L) = \sum_i 2^{1-i}\beta_i L^i$. With
\beq\label{WU}
	W=\Tr(w(L)),
\eeq
one generically looses the locality of $W$; the correct principle replacing it is that of pseudolocality, see \cite{IlievskietalQuasilocal,Doyon2017}. Here we simply assume $w(L)$ to have appropriate properties, see \cite{SpoToda} for a discussion.

The Gibbs state of the $N$-particle gas defined with respect to $H$, or more generally to $W$, does not exist, because the dynamical variables $x(m)$ tend to separate infinitely. As mentioned, one common way of defining finite-density states in the Toda model is to add a term proportional to the oriented volume $R$, which may be interpreted as a pressure $P$:
\beq
	W + PR.
\eeq
For any $P>0$ this confines the particles, and we may enquire about the (generalised) Gibbs free energy $G$ (with constant $N$, varying $R$) and associated state. We may instead consider states with varying number of particles while fixing the oriented volume, and add a chemical potential $\mu$,
\beq
	W-\mu N.
\eeq
There, we enquire about the (generalised) Landau potential $\Omega$ (with constant $R$, varying $N$) and associated state. In the chain viewpoint, $-\mu$ is instead interpreted as a pressure, as $N$ is the volume of the chain, and $-P$ is the chemical potential for the total topological charge $R$ (and the interpretations as Gibbs free energy and Landau potential is inverted).

We may then consider two (generalised) ensembles ${\cal E}_{\rm Gibbs}$ and ${\cal E}_{\rm Landau}$. In the first, we fix $N$ and let $R$ fluctuate, and set
\beq\label{gibbs}
	{\cal E}_{\rm Gibbs}\ :\ Z_{\rm Gibbs} = \int \prod_{m=1}^N \dd x(m) \dd p(m)\,\exp\big[-W - PR\big]
	\asymp \exp[-Ng]
\eeq
for large $N$. In the second, we fix $R$ and let $N$ fluctuate, and set
\beq\label{landau}
	{\cal E}_{\rm Landau}\ :\  Z_{\rm Landau} = \sum_{N=1}^{\infty}
	\int_{x(N)-x(1)=R} \prod_{m=1}^N \dd x(m) \dd p(m)\,\exp\big[-W + \mu N\big]
	\asymp \exp[-Rf]
\eeq
for large $R$. The gauge ambiguity under global shifts is fixed as per the discussion at the end of subsection \ref{ssecthamil}. Note that with respect to the standard definition of the thermodynamic pressure and chemical potential, we have absorbed the temperature (the Lagrange parameter $\beta_2$ associated to $H=Q_2$), as it does not play a special role in GGEs. We have also defined the specific Gibbs free energy $g=G/N$, and the specific (oriented) Landau potential $f = \Omega/R$.

Note from \eqref{gibbs} that $\p^2 g/ \p P^2<0$. Therefore, $\p g/\p P = \lim_{N\to\infty} \frc{\bra R\ket_{\bm\beta}}{N} = \nu$ is strictly decreasing with the pressure $P$. It is expected that for $P$ near to 0, $\nu\to\infty$ (as the system is unbounded), while as $P\to\infty$, $\nu\to-\infty$ (in particular the pressure is strong enough to change the ordering signature). As a consequence there is a single value of $P$ such that $\nu=0$, at which $g$ has a maximum.

By thermodynamic arguments, see below, we have
\beq\label{relation}
	g = \mu,\qquad f = - P.
\eeq
That is, at fixed $\beta_i$'s, the function $g$ of $-P$ given by \eqref{gibbs} is the inverse of the function $f$ of $\mu$ given by \eqref{landau}. Since $g$ has a maximum as a function of $P$, its inverse has two branches, determined by $\sgn(\nu)$. We may thus write
\beq
	g=g(f,\beta_1,\beta_2,\ldots),\qquad f = f(\sgn(\nu),g,\beta_1,\beta_2,\ldots).
\eeq

We in fact expect the two ensembles ${\cal E}_{\rm Gibbs}$ and ${\cal E}_{\rm Landau}$ to be equivalent: to reproduce the same averages and correlation functions of local observables. A point $\bm \beta$ in the manifold of state is therefore parametrised either by $f,\beta_1,\beta_2,\ldots$ or by $\sgn(\nu),g,\beta_1,\beta_2,\ldots$. That is, at fixed $\beta_i$'s and $\sgn(\nu)$, the ensemble is fixed by considering either $Q_0=N$, or $|Q_{\b 0}|=|R|$; the two charges are not set independently.

Relation \eqref{relation} is argued as follows. Consider the microcanonical partition function $Z_{\rm micro} = \int_{x(N)-x(1)=R} \prod_{m=1}^N \dd x(m) \dd p(m)\,\exp(-W)$. Then by the large-deviation principle applied to \eqref{landau}, $Z_{\rm micro} \asymp \re^{-RI_{\rm Landau}(\nu^{-1})}$ as $|R|\to\infty$ with fixed $\nu$, where $I_{\rm Landau}(\nu^{-1})$ is the Legendre transform of $f(\mu)$,
\beq\label{LTlandau}
	I_{\rm Landau}(\nu^{-1}) - f(\mu) = \mu\nu^{-1},\qquad I_{\rm Landau}'(\nu^{-1})=\mu.
\eeq
On the other hand, applied to \eqref{gibbs}, $Z_{\rm micro} \asymp \re^{-NI_{\rm Gibbs}(\nu)}$ as $N\to\infty$ with fixed $\nu$, where $I_{\rm Gibbs}(\nu)$ is the Legendre transform of $g(-P)$,
\beq
	I_{\rm Gibbs}(\nu) - g(-P) = -P\nu,\qquad I_{\rm Gibbs}'(\nu)=-P.
\eeq
Comparing the asymptotic expressions for $Z_{\rm micro}$, we deduce $\nu I_{\rm Landau}(\nu^{-1}) = I_{\rm Gibbs}(\nu)$. By inverse Legendre transform, this implies that $f(\mu)$ and $g(-P)$ are inverse of each other.

We may denote GGE averages by $\bra\cdots\ket_{\bm \beta}$. Below, it will be convenient to introduce a simplified notation for average densities and currents:
\beq
	\average{q}_i = \sgn(\nu)\bra\gas q_i\ket_{\bm \beta},\quad \average{j}_i = \sgn(\nu)\bra\gas j_i\ket_{\bm \beta},\quad
	\averagec{q}_i = \bra\chain q_i\ket_{\bm \beta},\quad \averagec{j}_i = \bra\chain j_i\ket_{\bm \beta}
\eeq
where the dependence on Lagrange parameters is implicit. In particular, we have introduced a factor $\sgn(\nu)$ for the gas densities and currents. This means that the density $\average{q}_i$ is a density {\em per unit oriented length}. This will simplify the expressions of thermodynamic and hydrodynamic quantities substantially; averages densities per unit oriented lengths have nicer properties as functions of $\nu$.

With \eqref{relation}, derivatives give rise to GGE averages: using \eqref{thermo}, \eqref{chargechain} and \eqref{chargegas}, we have
\beq\label{qchain}
	-\frc{\p g}{\p f} = \lim_{N\to\infty} \frc{\bra R\ket_{\bm\beta}}{N} = \nu=\averagec{q}_{\b 0},\qquad \frc{\p g}{\p\beta_i} = \lim_{N\to\infty}
	\frc{\bra Q_i\ket_{\bm \beta}}{N} = \averagec{q}_i
\eeq
and
\beq\label{qgas}
	-\frc{\p f}{\p g} = \lim_{|R|\to\infty} \frc{\bra N\ket_{\bm\beta}}{R} = \nu^{-1} = \average{q}_0,\qquad
	\frc{\p f}{\p\beta_i} = \lim_{|R|\to\infty}
	\frc{\bra Q_i\ket_{\bm \beta}}{R} = \average{q}_i.
\eeq
We note that $\averagec{q}_0 =\average{q}_{\b 0} = 1$. Clearly, the average conserved densities in the gas and the chain are related as
\beq\label{avchargechain}
	\averagec{q}_i= \nu \average{q}_i,\qquad i\in\{0,\b 0,1,2,3,\ldots\}.
\eeq

From this, when considering the gas viewpoint, it is more natural to think of the full set of conserved charges as those labelled by $\{0,1,2,3,\ldots\}$, while in the chain viewpoint, the more natural set is that labelled by $\{\b 0,1,2,3\ldots\}$.

\subsection{Hydrodynamics}\label{ssecthydro}

In the Euler-scale hydrodynamic description, an inhomogeneous, non-stationary state is assumed to be described as a collection of space-time dependent local fluid cells, each well approximated by GGEs (local entropy maximisation). Because the notion of space is different in the chain and in the gas, the emergent hydrodynamics is different. Importantly, fluid cells are naturally taken at fixed volume, with fluctuating number of particles or topological charge. The ensemble ${\cal E}_{\rm Landau}$ is therefore used for the gas hydrodynamics, while  ${\cal E}_{\rm Gibbs}$, for the chain hydrodynamics. In section \ref{sectghd} we derive the Euler hydrodynamic equations in both description. Here, we give the basic formulation of Euler-scale hydrodynamics in both cases.

For the chain, let a state be determined as in \eqref{gibbs} but with $W$ replaced by
\beq\label{Wellchain}
	\sum_{m\in\Z} \,\sum_{i\in\N} \beta_i(m/\ell)\chain{q}_i(m)
\eeq
and $PR$ replaced by
\beq
	\sum_{m\in\Z} P(m/\ell)r(m)
\eeq
for some $\ell>0$. Let us denote the averages by $\bra\dots\ket_\ell^{\rm ch}$. Then we set the Euler-scale fluid cell state $\bm\beta(y,t)$ by requiring that, for every local chain observable $\chain{o}(m,t)$, we have
\beq
	\bra \chain{o}\ket_{\bm\beta(y,t)} = \lim_{\ell\to\infty} \bra\chain{o}(\lfloor\ell y\rfloor,\ell t)\ket_\ell^{\rm ch}.
\eeq
Clearly $\bm\beta(y,0) = \bm \beta(y)$.  We denote by $\averagec{q}_i(y,t) = \bra \chain{q}_i\ket_{\bm\beta(y,t)}$ and $\averagec{j}_i(y,t) = \bra \chain{j}_i\ket_{\bm\beta(y,t)}$. From the conservation law \eqref{conschain},
\beq
	\p_t \averagec{q}_i(y,t)
	=  -\lim_{\ell\to\infty}
	\ell^{-1} \big(\averagec{j}_i(y+\ell^{-1},t) - \averagec{j}_i(y,t)\big)
	= -\p_y \averagec{j}_i(y,t)
\eeq
which are the Euler equations for the chain. In particular, with $\nu(y,t)$ defined by \eqref{qchain} and using \eqref{j0q1},
\beq
	\p_t \nu(y,t) = \p_y \averagec{q}_1(y,t).
\eeq

For the gas, let a state be determined as in \eqref{landau}, but with $W$ replaced by (compare with \eqref{Wellchain})
\beq
	\int_\R \dd x\, \,\sum_{i\in\N} \beta_i(x/\ell)\gas{q}_i(x)
\eeq
and $\mu N$ replaced by
\beq
	\int_\R \dd x\,\mu(x/\ell)\gas{q}_0(x)
\eeq
for some $\ell>0$. Let us denote averages by $\bra\dots\ket_\ell^{\rm ga}$. Then we set the Euler-scale fluid cell state $\bm\beta(x,t)$ by requiring that, for every local gas observable $\gas{o}(x,t)$, we have
\beq
	\bra \gas{o}\ket_{\bm\beta(x,t)} = \lim_{\ell\to\infty} \bra\gas{o}(\ell x,\ell t)\ket_\ell^{\rm ga}.
\eeq
Clearly $\bm \beta(x,0) = \bm \beta(x)$, and using the notation $\average{q}_i(x,t)$ and $\average{j}_i(x,t)$, we have
\beq\label{eulergas}
	\p_t \average{q}_i(x,t) + \p_x \average{j}_i(x,t) = 0,\qquad
	i\in\{0,1,2,3,\ldots\}.
\eeq

Note that, by convention, we use the variable $y$ for the scaled distances in the chain, and $x$ for the scaled distances in the gas.

\subsection{Relations for current averages}\label{ssectrelcur}

First, note that as follows from the above discussion,
\beq\label{j0q1}
	\averagec{j}_{\b 0} = -\nu\average{j}_{0} =
	-\nu\average{q}_1 = -\averagec{q}_1.
\eeq
Further, we have $\average{j}_{\b 0} = \averagec{j}_{0}=0$; the latter is immediate from the microscopic definition \eqref{def0}, while the former can be argued for by considering an inhomogeneous state at the Euler scale (see below), as in every local cell $\average{q}_{\b 0} =1$.

We now argue that
\beq\label{relcur}
	\averagec{j}_i = \average{j}_i
	- \averagec{q}_{1}\,\average{q}_i,\qquad i\in \N.
\eeq
This has the same structure as \eqref{curgas}, but it is at the level of GGE averages. This says that the chain current is obtained from the gas current by subtracting, from the gas current, the contribution coming from the motion of the particles themselves -- with average momentum density $\averagec{q}_{1}$. Equivalently, it says that we must {\em add} the quantity of charge transported by the topological current $\averagec{j}_{\b 0}$. These are, physically, the same effects. It is clear from the derivation below that this relation holds as well at the Euler scale of hydrodynamics, but it is not expected to hold at the diffusive scale.

Consider an Euler-scale hydrodynamic state as in subsection \ref{ssecthydro}. Let us assume that the state is such that
\beq\label{ass0}
	\averagec{q}_1(0,t) = 0
\eeq
for all $t$ in some finite interval. Then $x(0,t)$ stays around 0 for all $t$ of order $\ell$, up to fluctuations that are subleading in $\ell$, which we may assume to be of order $O(\ell^\alpha)$. Thus we can write
\beq
	x(\ell y,\ell t) = \sum_{m=0}^{\ell y} \nu(m/\ell, t) + O(\ell^\alpha) = \ell \int_0^{y} \dd y'\,\nu(y',t) + O(\ell^{\alpha}).
\eeq
From the definition of local gas observables, we also have
\beq
	\lim_{\ell\to\infty}\bra\gas{o}(\ell x,\ell t)\ket_\ell^{\rm ch} = \bra\gas{o}\ket_{\bm\beta(y,t)},\quad
	x = \int_0^{y} \dd y'\,\nu(y',t).
\eeq
Indeed, $\gas{o}(\ell x,\ell t)$ involves chain observables supported around the values of $m$ such that $\ell x =x(m,\ell t)$. These values of $m$ are given by $\ell \int_0^{y} \dd y'\,\nu(y',t)$ up to fluctuations which are of order $m^\alpha\sim \ell^\alpha$. In the Euler scaling limit, such fluctuations of the position give errors of the order $\ell^{\alpha-1}\to0$. Since the support is exponentially accurate, no additional power occurs from accumulation of these errors on the support. The conservation laws give
\beq
	\p_t \average{q}_i(y,t)|_x + \p_x\average{j}_i(y,t)|_t = 0.
\eeq

Changing variable,
\beq
	(\p_t y\,\p_y + \p_t)\average{q}_i(y,t)  + \p_xy \p_y\average{j}_i(y,t) =0.
\eeq
We have
\beq
	\p_x y = \nu(y,t)^{-1},\quad \p_t y = -\nu(y,t)^{-1}\int_0^y\dd y'\,\p_t\nu(y',t)=
	-\nu(y,t)^{-1}\averagec{q}_1(y,t)
\eeq
where we used \eqref{ass0}. So, omitting the explicit $y,t$ dependence for ligthness of notation, we have
\beqa
	\p_y \averagec{j}_i&=&
	-\p_t \averagec{q}_i \n &=&
	-\p_t (\nu \average{q}_i) \n &=&
	- \p_y \averagec{q}_1\, \average{q}_i
	-\nu\p_t  \average{q}_i \n &=&
	-\p_y \averagec{q}_1\, \average{q}_i
	-\nu \lt(-\p_xy \p_y\average{j}_i
	-\p_t y\,\p_y\average{q}_i\rt) \n &=&
	-\p_y \averagec{q}_1\, \average{q}_i
	+ \p_y\average{j}_i
	 - \averagec{q}_1\p_y\average{q}_i\n &=&
	\p_y \big(\average{j}_i-\averagec{q}_1 \average{q}_i
	\big).
\eeqa
Therefore, up to a $y$-independent term (an ambiguity in the definition of the current),
\beq
	\averagec{j}_i = \average{j}_i
	- \averagec{q}_1\,\average{q}_i.
\eeq

\subsection{Relations for hydrodynamic matrices}

In the context of Euler hydrodynamics, one defines a number of matrices which are useful in describing Euler-scale correlations \cite{Spohn-book,SpohnNonlinear,1742-5468-2015-3-P03007,SciPostPhys.3.6.039}. Here we discuss two fundamental matrices, the static covariance matrix $\magas{C}_{ij}$ and $\machain{C}_{ij}$, and the current susceptibility matrix $\magas{B}_{ij}$ and $\machain{B}_{ij}$, for the gas and the chain, respectively. Out of the static covariance and current susceptibility matrices, all other Euler-scale hydrodynamic correlations can be evaluated, including the Euler-scale dynamical correlation functions \cite{SpohnNonlinear,1742-5468-2015-3-P03007,SciPostPhys.3.6.039,doyoncorrelations}, the Drude weight \cite{SciPostPhys.3.6.039} and the transport large-deviation function \cite{Myers-Bhaseen-Harris-Doyon,DoMy19}.

Both matrices can be defined as space-integrated connected two-point correlation functions of local observables involving a conserved density, in homogeneous, stationary GGEs. The static covariance matrix in the gas is defined by
\beq\label{Cgas}
	\magas{C}_{ij} =  \sgn\nu\int_{\R} \dd x\,\big(\bra \gas{q}_i(x,0)\gas{q}_j(0,0)\ket_{\bm\beta} - 
	\bra \gas{q}_i(x,0)\ket_{\bm\beta} \bra\gas{q}_j(0,0)\ket_{\bm\beta} \big)
\eeq
for $i,j\in\{0,1,2,3,\ldots\}$. Note the presence of $\sgn(\nu)$ in our definition: this matrix can be seen as the integration over the {\em oriented} measure of a two-point function of {\em oriented} densities. Again, this non-standard definition confers to $\magas{C}_{ij}$ nicer properties. In the chain, it is defined similarly,
\beq
	\machain{C}_{ij} =  \sum_{m\in\Z} \,\big(\bra \chain{q}_i(m,0)\chain{q}_j(0,0)\ket_{\bm\beta} - 
	\bra \chain{q}_i(m,0)\ket_{\bm\beta} \bra\chain{q}_j(0,0)\ket_{\bm\beta} \big)
\eeq
for $i,j\in\{\b 0,1,2,3,\ldots\}$. This time, the derivative is taken at $f$ fixed. The current susceptibility matrices are defined by
\beq
	\magas{B}_{ij} =  \sgn\nu\int_{\R} \dd x\,\big(\bra \gas{q}_i(x,0)\gas{j}_j(0,0)\ket_{\bm\beta} - 
	\bra \gas{q}_i(x,0)\ket_{\bm\beta} \bra\gas{j}_j(0,0)\ket_{\bm\beta} \big)
\eeq
for $i,j\in\{0,1,2,3,\ldots\}$, and
\beq
	\machain{B}_{ij} =  \int \dd x\,\big(\bra \chain{q}_i(x,0)\chain{j}_j(0,0)\ket_{\bm\beta} - 
	\bra \chain{q}_i(x,0)\ket_{\bm\beta} \bra\chain{j}_j(0,0)\ket_{\bm\beta} \big)
\eeq
for $i,j\in\{\b 0,1,2,3,\ldots\}$.

It is clear that the static covariance matrices are symmetric. By general properties of homogeneous, stationary GGEs, the current susceptibility matrices also are \cite{SpohnNonlinear,PhysRevX.6.041065}: \beq\label{symmetry}
	\magas{C}_{ij} = \magas{C}_{ji},\quad
	\machain{C}_{ij} = \machain{C}_{ji},\quad
	\magas{B}_{ij} = \magas{B}_{ji},\quad
	\machain{B}_{ij} = \machain{B}_{ji}.
\eeq

As is clear from the definition of the ensembles ${\cal E}_{\rm Gibbs}$ and ${\cal E}_{\rm Landau}$ in \eqref{gibbs} and \eqref{landau}, the static covariance and current susceptibility matrices for indices $i,j\in\N$ can be obtained by differentiation of averages of conserved densities and currents with respect to the Lagrange parameters $\beta_i$'s. Since the ensembles ${\cal E}_{\rm Gibbs}$ and ${\cal E}_{\rm Landau}$ are expected to be equivalent, they reproduce, after taking the thermodynamic limit, the same connected correlation  functions at fixed point, which decay exponentially and thus are integrable. However, susceptibilities, variations of averages with respect to Lagrange parameters, are ensemble-dependent. Indeed, the order of limit is important: the thermodynamic limit of space-integrated connected correlation functions depends on the ensemble chosen. One must choose the ensemble where the volume is fixed, avoiding large volume fluctuations which may contribute to the limit. For the gas, with volume $|R|$, this is ${\cal E}_{\rm Landau}$, while for the chain, with volume $N$, this is ${\cal E}_{\rm Gibbs}$. Within ${\cal E}_{\rm Gibbs}$, one takes derivatives at fixed $f$, while within ${\cal E}_{\rm Landau}$, it is at fixed $g$. Thus, we have
\beq\label{Cgasd}
	\magas{C}_{ij} \stackrel{i\neq 0}=
	-\frc{\p \average{q}_j}{\p\beta_i}\Big|_g,\quad
	\magas{C}_{00} = \frc{\p \average{q}_0}{\p g},\quad
	\machain{C}_{ij} \stackrel{i\neq \b 0}=
	-\frc{\p \averagec{q}_j}{\p\beta_i}\Big|_f,\quad
	\machain{C}_{\b 0 \b 0} = \frc{\p\averagec{q}_{\b 0}}{\p f}
\eeq
and
\beq
	\magas{B}_{ij}  \stackrel{i\neq  0}=
	-\frc{\p \average{j}_j}{\p\beta_i}\Big|_g,\quad
	\magas{B}_{00} = \frc{\p \average{j}_0}{\p g},\quad
	\machain{B}_{ij} \stackrel{i\neq \b 0}=
	-\frc{\p \averagec{j}_j}{\p\beta_i}\Big|_f,\quad
	\machain{B}_{\b 0 \b 0} = \frc{\p\averagec{j}_{\b 0}}{\p f}.
\eeq

We now establish the relations between gas and chain hydrodynamic matrices. First, using \eqref{qchain}, \eqref{qgas} and the chain rule, we have
\beq\label{C00}
	\machain{C}_{\b 0 \b 0} = \frc{\p \nu}{\p f} =
	\frc{\p}{\p f}\frc1{\average{q}_{0}} = \frc{\p g}{\p f} \frc{\p}{\p g}\frc1{\average{q}_0}
	=\nu^3 \magas{C}_{00}.
\eeq
Second, for $i\in\N$,
\beq
	\machain{C}_{i \b 0} = -\frc{\p \nu}{\p\beta_i}\Big|_f =
	-\frc{\p }{\p\beta_i}\frc1{\average{q}_0}\Big|_f
	= \nu^2\lt(-\magas{C}_{i0} + \frc{\p g}{\p\beta_i}\Big|_f \frc{\p\average{q}_0}{\p g}\rt)
	= -\nu^2\lt(\magas{C}_{i0} - \averagec{q}_i \magas{C}_{00}\rt).
\eeq
Then, using \eqref{avchargechain} and the symmetric of $\magas{C}_{ij}$, for $i,j\in\N$,
\beq
	\machain{C}_{ij} = -\frc{\p}{\p\beta_i} (\nu\average{q}_j)\Big|_f=
	\machain{C}_{i\b 0}\average{q}_j + \nu (\magas{C}_{ij} -  \averagec{q}_i \magas{C}_{j0})
	= \nu\big(\magas{C}_{ij} 
	-\averagec{q}_j \magas{C}_{i0} - 
	\averagec{q}_i \magas{C}_{j0} + \averagec{q}_i\averagec{q}_j\magas{C}_{00}\big).
\eeq
Note that these expressions respect the symmetry relations \eqref{symmetry}.

Similar calculations can be done for the current susceptibility matrices, using \eqref{j0q1} and \eqref{relcur}. The results are, again for $i,j\in\N$:
\beqa
	\machain{B}_{\b 0 \b 0} &=& -\machain{C}_{\b 0 1} 
	\\
	\machain{B}_{i \b 0} &=& -\machain{C}_{i1} 
	\\
	\machain{B}_{ij} &=& \magas{B}_{ij} - \averagec{q}_i \magas{C}_{j1} - \averagec{q}_j \magas{C}_{i1}
	+ \averagec{q}_i\averagec{q}_j \magas{C}_{01}
	+ \averagec{q}_1\averagec{q}_i \magas{C}_{j0} + \averagec{q}_1\averagec{q}_j \magas{C}_{i0}
	-  \averagec{q}_i \averagec{q}_j \averagec{q}_1\magas{C}_{00}. \label{Bij}
\eeqa

\section{Generalised hydrodynamics of the Toda system}\label{sectghd}

We now develop the generalised hydrodynamics (GHD) of the Toda model. For this purpose, we need to express the equations of state of the model: the relation between the currents and the conserved densities. We use the quasiparticle description of the thermodynamics, based on the TBA, and in particular the prescriptions that extend its application to classical models \cite{10.21468/SciPostPhys.4.6.045}. Because of relation \eqref{relcur}, it is sufficient to do this either for the Toda gas, or the Toda chain; once one is done, the results for the other follow. Here we choose the Toda gas, as it has much simpler asymptotic states, hence simpler TBA. We obtain the TBA and GHD by evaluating the exact classical scattering shift of the quasiparticles of the asymptotic states. The quasiparticles are simply the velocity tracers of the particles of the gas itself, in parallel to the situation for the hard rod gas \cite{DYC18}.

In thermal states, we find agreement with the ``semi-classical" derivation of Theodorakopoulos \cite{T84} and Opper \cite{O85}. This derivation was based on taking a semi-classical limit of a quantum Bethe ansatz description. We also find agreement, for the thermodynamics in the full GGE, with the more rigorous derivation based on the relation between the Toda {\em chain} and random matrix theory \cite{SpoToda}.

Our main results are the exact equations of state, both in the gas and in the chain, and the hydrodynamic equations. Since these are expressed within the GHD formalism, this in principle leads to many exact predictions, including the exact profile for domain-wall initial conditions both for gas and chain observables, and Euler-scale correlations, transport fluctuations and diffusive corrections in the gas.

\subsection{Asymptotic states and scattering}\label{ssectscat}

Consider a configuration of the Toda gas with finite $N$. This can be seen as the result of an infinite-time evolution from a configuration of particles infinitely separated, and at appropriate momenta so that the extrapolated trajectories converge towards a finite interval in $\R$. The description of this asymptotic configuration is the representation of the original gas configuration as an asymptotic in-state. Likewise, the original gas configuration tends, at large times, to one with infinitely separated particles, an out-state. We can therefore characterise configurations by scattering states. Trajectories have the form
\beq
	x(m,t) = p_m^{\rm in} t + x_m^{\rm in} + O(t^{-\infty}),\quad p(m,t) = p_m^{\rm in} + O(t^{-\infty})\qquad (t\to-\infty)
\eeq
and
\beq
	x(m,t) = p_m^{\rm out} t + x_m^{\rm out} + O(t^{-\infty}),\quad p(m,t) = p_m^{\rm in} + O(t^{-\infty})\qquad (t\to+\infty).
\eeq
The $p_m^{\rm in,out}$'s are the asymptotic momenta, and the $x_m^{\rm in,out}$'s are the ``impact parameters".

The asymptotic momenta and impact parameters characterise completely the gas's configuration at time 0, and in fact form a canonical system of coordinates whose time evolution is that of free particles. Indeed, defining the scattering map
\[
	S = \lim_{t\to-\infty} \re^{t\{\cdot,H\}}\re^{-t\{\cdot,H_0\}} = \lim_{t\to-\infty} \re^{-t\{\cdot,\t H_0\}}\re^{t\{\cdot,H\}}
\]
where $H_0 = \sum_m p(m)^2/2$ is the free Hamiltonian on time-zero coordinates and $\t H_0 = \sum_m p_m^2/2$ is the free Hamiltonian on asymptotic coordinates, we have $p_m^{\rm in} = S(p(m))$ and $x_m^{\rm in} = S(x(m))$, as well as $S^{-1}(H) = H_0$.

In the Toda gas, the scattering of particles is elastic and factorizable into two-particle scattering processes.  That is, $\{p_m^{\rm in}\} = \{p_m^{\rm out}\}$, and $x_m^{\rm in} = x_{N-m+1}^{\rm out} + \Delta_m$, where the shifts $\Delta_m$ can be evaluated by assuming that the many-body scattering has occurred as a sequence of well separated two-body scattering events (no matter the order in which these occur). The asymptotic momenta may be seen as the ``action" variables, and the impact parameters as the ``angle" variables, in the action-angle description of integrable systems.

The statements in the above paragraph can be shown rigorously, see \cite{Mo75}. Here for completeness we give a sketch of a perhaps simpler proof that follows arguments used in quantum field theory \cite{Parke80}. This also helps  defining the concept of quasiparticle in finite-density gases, which we will make use of. See also Section \ref{secttba} for a description of how the impact parameters relate to the particles' positions at time 0 when scattering is elastic and factorizable, and how they are compactified for gases in finite volume.

The scattering form is immediate from the repulsive nature of the interaction. Consider the conserved charge $Q_3$. In order to evaluate the out momenta and impact parameters, we act with the Poisson flow generated by $Q_3$ for a flow time $t_3$ at a large negative time $-T$, and for a flow time $-t_3$ at the large positive time $T>0$. This does not affect the resulting gas configuration at times $t>T$, as the $Q_3$ flow commutes with that generated by the hamiltonian. On the well separated particles at large $|t|$, $Q_3$ acts as $\sum_m p(m,t)^3/4$ up to exponentially small corrections due to the terms $\re^{-r(m,t)}$. This action can be evaluated:
\beq\begin{aligned}
	\{x(m,t),Q_3\} &= \frc14 \sum_{m'=1}^n \{x(m,t),p(m',t)^3\} + O(t^{-\infty})\\ & = \frc34 p(m,t)^2 + O(t^{-\infty}) =
	\frc34 (p_m^{\rm in,out})^2 + O(t^{-\infty}).
	\end{aligned}
\eeq
That is, the flow shifts the impact parameters by quantities that are nonlinear in momenta, and does not affect the momenta themselves. As particles move, in time, by distances which grow linearly with the momenta, it is simple to see that sufficiently far along the flow (for $t_3$ large enough), the scattering of particles will occur as a series of two-body scattering processes that are well separated in space-time. Thanks to the exponentially decaying strength of the force between particles, these scattering events can be considered separately, and thus scattering is factorised.  Since in one dimension a two-body scattering event preserves both momenta, scattering is also elastic.

Note that the above argument holds except for a measure-zero subset in the space of asymptotic momenta: for certain momenta, three or more particles may still meet within a small space-time region for all $t_3$. The argument may be completed by assuming, or showing, continuity of the scattering shifts $\Delta_m$ as functions of the momenta. One may alternatively consider higher conserved charges $Q_4, Q_5,\ldots$. As more conserved charges are considered, the dimension of the subset of asymptotic momenta, where collisions happen in finite space-time regions for all $t_3,t_4,t_5,\ldots$, diminishes.

It is a simple matter to work out the exact scattering position shift $\varphi(p_1-p_2)$ that affects particles in a two-body process with in-momenta $p_1$ and $p_2$. In such a process, particles exchange their momenta, as they cannot cross. By convention, if the extrapolated trajectory of the left-most particle from the far past lies to the {\em right} of the trajectory of the right-most particle going far in the future -- that is, there has been a {\em retreat} of the carrier of the greater momentum, or a time {\em delay} -- then the shift of trajectory is the positive quantity $\varphi(p_1-p_2)>0$; otherwise, it is $\varphi(p_1-p_2)<0$

In order to evaluate $\varphi(p_1-p_2)$, we consider a Toda gas composed of only two particles, $N=2$. By Galilean invariance, we may assume that the two asymptotic in-momenta are $p_1=p$ and $p_2=-p$, with $p>0$. By parity symmetry of the evolution equation, we then have $x(1,t) = -x(2,t) = x(t)$ as well as $p(1,t)=-p(2,t)=p(t)$, and we set $x(t) = pt+o(t)$ and $p(t)= p +o(1)$ as $t\to-\infty$ for the asymptotic in-state condition.  That is, the left and right in-asymptotes are the lines of slopes $p$ and $-p$, respectively, which cross at time 0. By time-reversal symmetry, the trajectories are symmetric with respect to some time $t_0$. Clearly, $x(t)$ increases with time until it reaches its maximum, at some time $t_0$ at which $p(t_0)=0$, and then reverses its direction and decreases towards negative infinity at infinite time, with $x(t)= -pt+O(1)$ as $t\to\infty$. Therefore, the trajectory of $x(t)$ at large times asymptotes to the line of slope $-p$ which crosses the left in-asymptote exactly at $t_0$. That is, the out-asymptotes are the lines of slope $-p$ and $p$ which cross, respectively, the left and right in-asymptotes at time $t_0$. The scattering shift is therefore
\beq
	\varphi(2p) = 2pt_0.
\eeq
The conserved energy is
\beq
	H = p(t)^2 + \re^{2x(t)} = p^2
\eeq
where the last equation is from evaluating it as $t\to-\infty$. Therefore
\beq
	p(t) = p - \frc{\re^{-2pt}}{2p} + O(\re^{-4pt}),\qquad t\to-\infty.
\eeq
The equations of motion give
\beq
	\frc{\dd p(t)}{\dd t} = \re^{2x(t)} = p^2-p(t)^2.
\eeq
The time $t_0$ is evaluated by taking, in the limit $T$ large,
\beq
	T + t_0 = \int_{-T}^{t_0}\dd t
	= \int_{-T}^{t_0}\frc{\dd p(t)}{\dd p(t)/\dd t} = \int_{p - \frc{\re^{-2pt}}{2p}}^{0}\frc{\dd q}{p^2-q^2}
	+o(1) = T + \frc{\log 2p}{p} + o(1).
\eeq
Therefore, $\varphi(2p) = 2\log 2p$. We extend the function to all $p\in\R$ by symmetry,
\beq\label{shift}
	\varphi(p) = 2\log |p|.
\eeq

Finally, consider a many-particle scattering process, leading to a finite-density gas. Here, let us assume that the overall ordering agrees with the label ordering (positive ordering signature). Since momenta are preserved, it is convenient to define a {\em quasiparticle} as the tracer of a momentum. The quasiparticle jumps amongst actual particles upon collisions. An argument using $Q_3$ as above allows one to disentangle all collisions, thus to follow a quasiparticle. The $Q_3$ action on a particle is affected by nearby particles in a way that decays exponentially with the distance. Hence, quasiparticles have well defined positions up to corrections due to nearby particles; the strength of such corrections decays exponentially with the distance. In finite-density gases, quasiparticle positions are therefore precise up to order-1 corrections. These positions are affected by scattering shifts at each collision. This quasiparticle concept will be useful below.

\begin{rema}
As mentioned, in the chain viewpoint, the asymptotic states take a different form. In this case, the volume of the chain $N$ is taken large, and we consider the asymptotic forms of states with finite $R$, at large positive or negative times. The asymptotic form is known to be decomposable into radiative corrections on top of a set of solitons \cite{ZK65,KT}. The scattering properties of these objects is more complicated and has been worked out, with proposals for the associated thermodynamics along the lines of TBA \cite{TI86}. Here we do not discuss these aspects.
\end{rema}

\subsection{Classical thermodynamic Bethe ansatz for the gas}

In an $N$-particle scattering state with asymptotic momenta $p_m$, asymptotically in time, the Lax matrix $L(t)$ is diagonal with eigenvalues $p_m$, as the off-diagonal elements tend exponentially fast to zero, $\re^{-r(m,t)/2}\to 0$.  Effectively, the Toda time evolution diagonalises the Lax matrix, and the asymptotic state's momenta are (proportional to) the eigenvalues. Since $W$ (see \eqref{WU}) is conserved in time, in such a state we have
\beq\label{Was}
	W = \sum_{m=1}^N w(p_m).
\eeq
In particular, the conserved charges $Q_i$ take the form
\beq\label{Qi}
	Q_i = \sum_{m=1}^N h_i(p)
\eeq
where
\beq
	h_i(p) = \frc{p^i}{2^{i-1}}\quad (i\in\N),\qquad
	w(p) = \sum_{i=1}^{\infty}\beta_i h_i(p).
\eeq
We also define $h_0(p)=1$.

From the knowledge of the scattering shift and of the conserved charges in scattering states, the TBA technology immediately gives a conjecture for the exact specific Landau potential $f$ in \eqref{landau} for any GGE, in terms of $\beta_i$'s and $g$. The proposed expression, for classical particle systems, is 
\beq\label{f}
	f = -\int \dd p\,\re^{-\ep(p)}
\eeq
where the pseudoenergy solves
\beq\label{ep}
	\ep(p) = w(p) - g - \int \dd p'\,\varphi(p-p')\,\re^{-\ep(p')}.
\eeq
As per the discussion in subsection \ref{ssectgge}, the free energies is determined by $g$ and $\sgn(\nu)$. Accordingly, given $w(p)$ and $g$ (smaller than its maximal value), there are two solutions to \eqref{ep}, which, in the equations below, are distinguished by $\sgn(\nu)$.

We provide a derivation of \eqref{f} with \eqref{ep}, for positive ordering signature $\sgn(\nu)=+$, in  section \ref{secttba}, purely form the definition of the specific Landau potential and the classical factorised scattering.

From \eqref{f} and \eqref{ep}, it is possible to obtain expressions for the conserved densities. One defines the occupation function \cite{10.21468/SciPostPhys.4.6.045}
\beq
	n(p) = \re^{-\ep(p)},
\eeq
and the dressing operation of a function $h(p)$, as the solution to
\beq\label{dressing}
	h^{\rm dr}(p) = h(p) + \int \dd p'\,\varphi(p-p')\, n(p')h^{\rm dr}(p').
\eeq
From this, using $\p\ep(p)/\p\beta_i = h_i^{\rm dr}(p)$, by differentiation one obtains
\beq\label{qidress}
	\average{q}_i = \int \dd p\,n(p)h_i^{\rm dr}(p).
\eeq
This is valid for all $i\in\{0,1,2,\ldots\}$, and in particular
\beq\label{num1}
	\nu^{-1} = \int \dd p\,n(p)1^{\rm dr}(p)
\eeq
where $1^{\rm dr}(p)$ is the dressing of the function $1(p)=1$. With the useful identity
\beq\label{id}
	\int \dd p\,h^{\rm dr}(p)n(p)\t h(p) = \int \dd p\,h(p)n(p)\t h^{\rm dr}(p),
\eeq
we may also write
\beq\label{avgas}
	\average{q}_i = \int \dd p\,\rho_{\rm p}(p)h_i(p) ,\qquad
	i\in\{0,1,2,3,\ldots\}
\eeq
with
\beq\label{rhoprhos}
	\rho_{\rm p}(p) = 1^{\rm dr}(p)\,n(p) = \rho_{\rm s}(p) n(p),
\eeq
where $\rho_{\rm s}(p) = 1^{\rm dr}(p)$ is a ``state density", given, using \eqref{dressing}, by
\beq\label{statedensity}
	\rho_{\rm s}(p) = 1 + 2\int \dd p'\,\log|p-p'|\,\rho_{\rm p}(p').
\eeq
In particular,
\beq\label{rhonu}
	\int \dd p\,\rho_{\rm p}(p) = \nu^{-1}.
\eeq

Equations \eqref{avgas} and \eqref{rhonu} have a natural interpretation. With \eqref{Qi}, we may see $\rho_{\rm p}(p)$ as a {\em density of quasiparticles} with momenta $p\in\R$ per unit oriented gas phase space element. That is, $\sgn(\nu) \dd p \dd x\rho_{\rm p}(p)$ is the number of quasiparticles with momenta $p$ in the phase-space element $\dd p \dd x$ (in the case of negative ordering signature or with $\nu=0$, this interpretation, as well as a the behaviour of the various TBA quantities above, would need more investigation).   This leads to the quasiparticle interpretation of GGEs. In this interpretation, a distribution of configurations, as controlled by the Lagrange parameters $\bm\beta$, can be formulated as a distribution of asymptotic states, either in or out.  According to the usual microcanonical-macrocanonical correspondence of thermodynamics, it is sufficient, for the description of a thermodynamic state, to consider a fixed set of asymptotic particles' momenta. We then parametrise the GGEs $\bm\beta$ by densities $\rho_{\rm p}(p)$,
\beq\label{relbeta}
	\bm\beta\ \leftrightarrow \ \rho_{\rm p}(p).
\eeq
Since momenta are conserved quantities, the set of all GGEs corresponds to an infinite-dimensional space of functions $\rho_{\rm p}(\theta)$.

\subsection{Generalised hydrodynamics for the gas}

With the above TBA formulation, we can immediately borrow the results and arguments developed recently in the context of GHD.

The exact currents can be evaluated in terms of $\rho_{\rm p}(p)$. A simple argument, based on the quasiparticle interpretation of GGEs, is to modify the velocity $p$ of a test quasiparticle by adding the jumps $\varphi(p-p')$ made as it travels within a gas of other quasiparticles of momenta $p'$ distributed according to $\rho_{\rm p}(p')$. This argument has been used for instance for soliton gases \cite{Zakharov,El-2003,El-Kamchatnov-2005,El2011}, and made generally in \cite{DYC18} in the context of GHD; it only necessitates the knowledge of the jump function $\varphi(p)$. The result is that the ``bare" velocity $v(p)=p$ of the particles is modified in a way that depends on the state and on the jump function $\varphi(p)$. This leads to an ``effective" velocity, defined as the solution to a linear integral equation which here takes the form
\beq\label{veff}
	v^{\rm eff}(p) = p + 2\int \dd p'\,\log|p-p'|\, \rho_{\rm p}(p')(v^{\rm eff}(p')-v^{\rm eff}(p)).
\eeq
The result for the currents, involving the effective velocity, is then
\beq\label{jieff}
	\average{j}_i = \int \dd p\,v^{\rm eff}(p) \rho_{\rm p}(p) h_i(p).
\eeq
With this, the Euler hydrodynamic equations \eqref{eulergas} give, as usual with the assumption of completeness of the set of conserved quantities \cite{PhysRevX.6.041065,PhysRevLett.117.207201},
\beq\label{eulergas}
	\p_t \rho_{\rm p}(p;x,t) + \p_x\big(v^{\rm eff}(p;x,t) \rho_{\rm p}(p;x,t)\big) = 0
\eeq
where the state and the corresponding effective velocity depend on the space-time position of the fluid cell.

Note that the argument leading to \eqref{veff} and \eqref{jieff} usually assumes $\rho_{\rm p}(p)>0$. However, having chosen to consider oriented densities, we expect these expressions to hold for any ordering signature. We will argue {\em a posteriori} that these lead to expressions for chain currents that are well behaved -- supposedly analytic -- in $\nu$ around $\nu=0$, as it should.

From \eqref{veff} and \eqref{dressing} one can also show that \cite{PhysRevX.6.041065,PhysRevLett.117.207201}
\beq\label{veffdr}
	v^{\rm eff}(p) = \frc{v^{\rm dr}(p)}{1^{\rm dr}(p)}.
\eeq
Therefore, using the velocity $v(p)=p$ and \eqref{id}, we have
\beq
	\average{j}_i = \int \dd p\,p\,n(p)h_i^{\rm dr}(p).
\eeq
In terms of the occupation function, the Euler equations \eqref{eulergas} take the form \cite{PhysRevX.6.041065,PhysRevLett.117.207201}
\beq
	\p_t n(p;x,t) + v^{\rm eff}(p;x,t)\p_x n(p;x,t)=0.
\eeq
That is, the occupation function form the normal coordinates of the fluid.

The various hydrodynamic operators, Euler-scale correlation functions, solution to the partitioning protocol, and diffusion operator can then immediately be read off from results in the literature \cite{SciPostPhys.3.6.039,PhysRevX.6.041065,PhysRevLett.117.207201,dNBD,dNBD2}. In particular, we quote \cite{SciPostPhys.3.6.039}
\beqa
	\magas{C}_{ij} &=& \int \dd p\,\rho_{\rm p}(p)h_i^{\rm dr}(p)h_j^{\rm dr}(p) \n
	\magas{B}_{ij} &=& \int \dd p\,\rho_{\rm p}(p)v^{\rm eff}(p) h_i^{\rm dr}(p)h_j^{\rm dr}(p).
	\label{CB}
\eeqa

\subsection{Generalised thermodynamics and hydrodynamics for the chain}

For the chain thermodynamics, we may directly use \eqref{avchargechain}, and we find
\beq\label{avchain}
	\averagec{q}_i = \nu \int \dd p\,\rho_{\rm p}(p)h_i(p) 
	= \nu \int \dd p\, n(p) h_i^{\rm dr}(p)\quad (i\in\N),\qquad
	\averagec{q}_{\b 0} = \nu = \frc{1}{\int \dd p\, \rho_{\rm p}(p)}
\eeq
(in fact, the relations for $\averagec{q}_i$ also hold at $i=0$, giving $\averagec{q}_0=1$).  This can also be deduced from the specific Gibbs free energy $g$ defined by \eqref{ep} and constraint \eqref{f}, by taking derivatives as in \eqref{qchain}. Differentiating \eqref{ep} we indeed obtain the same results; for instance, for $i\in\N$,
\beq
	\frc{\p \ep(p)}{\p\beta_i} = (h_i-\averagec{q}_i)^{\rm dr}(p) = h_i^{\rm dr}(p) - \averagec{q}_i \rho_{\rm s}(p)
\eeq
and the constraint \eqref{f} gives, differentiating at $f$ fixed,
\beq
	0 = \int \dd p\,\frc{\p \ep(p)}{\p\beta_i} n(p) = 
	\int \dd p\,h_i^{\rm dr}(p) n(p) - \averagec{q}_i \int \dd p\,\rho_{\rm p}(p).
\eeq

Interestingly, it is possible to express these results in a simpler-looking fashion, where the dressing operation and the denominator are both accounted for by a derivative with respect to $f=-P$. Indeed, from \eqref{qchain} and \eqref{ep} we have
\beq
	\frc{\p\ep(p)}{\p f} = -\lt(\frc{\p g}{\p f}\rt)^{\rm dr}(p) = -\frc{\p g}{\p f} \rho_{\rm s}(p) = \nu \rho_{\rm s}(p)
\eeq
and therefore
\beq
	\frc{\p n(p)}{\p f}= -\nu \rho_{\rm p}(p).
\eeq
That is, the pressure derivative of the gas occupation function is the quasiparticle density per unit phase-space  with respect to lengths on the chain; it is a density, instead of an oriented density, and we would expect $\nu \rho_{\rm p}(p)>0$. This allows us to write
\beq
	\averagec{q}_i = -\frc{\p}{\p f}\int \dd p\,n(p) h_i(p)\quad (i\in\N).
\eeq
These expressions of chain density averages in terms of pressure derivatives are obtained from random matrix theory in \cite{SpoToda}, therefore showing the validity of the general TBA formalism in the present case.

From \eqref{j0q1} and \eqref{relcur} we also obtain expressions for all chain currents. Explicitly, in various equivalent forms, the current of the charge $Q_{\b 0}$ is
\beq\label{j0chain}
	\averagec{j}_{\b 0} = - \nu \int \dd p\,p\rho_{\rm p}(p) = -\nu \int \dd p\,v^{\rm dr}(p)n(p)=
	\frc{\p}{\p f}\int \dd p\,p\,n(p)
\eeq
and, for $i\in\N$, we find
\beq\label{jichain}
	\averagec{j}_i 
	= \int \dd p\,(v^{\rm eff}(p)-\averagec{q}_1)\rho_{\rm p}(p)h_i(p)
	= \int \dd p\,(p-\averagec{q}_1)n(p) h_i^{\rm dr}(p)
\eeq
(these expressions hold also for $i=0$, giving $\averagec{j}_0=0$).

From this we can pass to the Euler scale and obtain the Euler hydrodynamic equations. The Euler scale is reached by going to the $y$ variable as explained in subsection \ref{ssectrelcur}. The Euler equations are written separately for the quasiparticle density $\rho_{\rm p}(p;y,t)$ and the density of topological charge $\nu(y,t)$. We assume a complete set of functions $h_i(p)$ is used, and obtain
\beq\label{eulerchain}
	\p_t \big[\nu(y,t) \rho_{\rm p}(p;y,t)\big] + \p_y\big[(v^{\rm eff}(p;y,t)-\averagec{q}_1(y,t))\rho_{\rm p}(p;y,t)\big] = 0
\eeq
and
\beq\label{eulerchainnu}
	\p_t \nu(y,t) = \p_y \averagec{q}_1(y,t).
\eeq
Note that the first equation preserves the normalisaion
\beq\label{norm}
	\nu\int \dd p\,\rho_{\rm p}(p;y,t) = 1,
\eeq
and thus from a naive counting, the number of equations is the same as in the case of the gas.

We show below that the occupation function $n(p;y,t)$ still normalises the equation \eqref{eulerchain},
\beq\label{chainnormal} 
	\nu \p_tn(p;y,t) +  \big[v^{\rm eff}(p;y,t)- \averagec{q}_1(y,t)\big]\p_y n(p;y,t) = 0.
\eeq
One can check that this equation, along with the second set of equations in \eqref{avchain} for the expression of $\nu$, implies \eqref{eulerchainnu}; hence it forms a complete set or Euler equations.

The above formulae suggest the definition of the chain quasiparticle density
\beq
	\rho_{\rm p}^{\rm c}(p) = \nu \rho_{\rm p}(p)
\eeq
and of the chain effective velocity
\beq\label{veffchain}
	v^{\rm eff,  c}(p) =  \nu^{-1}\big(v^{\rm eff}(p) - \averagec{q}_1\big).
\eeq
As mentioned, the former is clearly a quasiparticle density per unit phase-space  with respect to lengths on the chain. The latter is the effective velocity of gas quasiparticles with respect to the frame where the chain momentum is zero, and rescaled by the oriented gas particle density. Going to the zero-momentum frame subtract the contribution to the gas effective velocities that comes from the motion of the particles themselves, keeping only that corresponding to the transport of conserved quantities from site to site. The factor $|\nu|^{-1}$ rescales the distance appropriately in order to have velocities on the chain, and the ordering signature $\sgn(\nu)$ changes, if need be, the direction of transport so that it be with respect to the chain orientation.

The chain effective velocity also has a ``microscopic" underpinning in terms of quasiparticles, much like for the gas effective velocity. Indeed, in the chain, the effective velocity for the transfer of the conserved charge corresponding to the quasiparticle of momentum $p$, can be thought of as the oriented number of crossing this quasiparticle makes with other quasiparticles in a period of time, divided by the time. This is because every crossing corresponds to the passage of the conserved charge from one particle to another, hence from one site to another; thus a displacement of one unit in chain lengths. By the standard arguments, the oriented number of crossing is the integral over all other quasiparticles $q$ of the density $\rho_{\rm p}(q)$ of quasiparticles, times the probability $|v^{\rm eff}(p)-v^{\rm eff}(q)|$ and the direction $\sgn(v^{\rm eff}(p)-v^{\rm eff}(q))$. Thus we may write
\beq
	v^{\rm eff, c} = \int \dd q\,\rho_{\rm p}(q)(v^{\rm eff}(p)-v^{\rm eff}(q))
	= \nu^{-1}v^{\rm eff}(p) - \average{j}_0
	= \nu^{-1}(v^{\rm eff}(p) - \averagec{q}_1)
\eeq
in agreement with \eqref{veffchain}.

These quantities are expected to be well behaved in $\nu$ around $\nu=0$. Indeed, the gas oriented density $\rho_{\rm p}(p)$ may diverge, but multiplying by $\nu$, the result is expected to be finite. Similarly, since at $\nu=0$ a macroscopic number of gas particles accumulate around a small volume, all effective velocities must take values very near to the overall momentum $\averagec{q}_1$ of this macroscopic number of particles. This is expected to lead to analyticity at $\nu=0$ of the average charges and currents.

The average charges and currents now take the form
\beq
	\averagec{q}_i = \int \dd p\,\rho_{\rm p}^{\rm c}(p)h_i(p),\qquad
	\averagec{j}_i 
	= \int \dd p\,v^{\rm eff, c}(p)\rho_{\rm p}^{\rm c}(p)h_i(p).
\eeq
The Euler equations take the simple form
\beq\label{eulerchainform}
	\p_t \rho_{\rm p}^{\rm c}(p;y,t) + \p_y\big(v^{\rm eff, c}(p;y,t)\rho_{\rm p}^{\rm c}(p;y,t)\big) = 0
\eeq
with \eqref{eulerchainnu}, the constraint
\beq
	\int \dd p\,\rho_{\rm p}^{\rm c}(p;y,t) = 1
\eeq
being automatically preserved. Equivalently, in normal form it is the single equation
\beq
	\p_tn(p;y,t) + v^{\rm eff, c}(p;y,t)\p_y n(p;y,t) = 0.
\eeq
The effective chain velocity $v^{\rm eff, c}(p;y,t)$ can be expressed solely as a functional of $n(\cdot;y,t)$ using \eqref{avchain} and \eqref{veffdr} along with \eqref{rhoprhos} and \eqref{dressing}. See Appendix \ref{app} for a discussion of the relation with the usual GHD structure.

Finally, relations \eqref{C00}-\eqref{Bij} along with the results  \eqref{CB} for hydrodynamic matrices in the gas give expressions for those in the chain (here written in a compact notation):
\beqa
	\machain{C}_{\b 0 \b 0} &=& \nu^3 \int \dd p\,\rho_{\rm p} 1^{\rm dr}1^{\rm dr}\n
	\machain{C}_{i \b 0} &=& -\nu^2 \int \dd p\,\dd \t p\,h_i^{\rm dr} {\t 1}^{\rm dr}
	\lt(\rho_{\rm p}\delta(p-\t p) - \nu n \t \rho_{\rm s}\rt)\n
	\machain{C}_{ij} &=& \nu\int \dd p\,\dd \t p\,h_i^{\rm dr} {\t h}_j^{\rm dr}
	\lt(\rho_{\rm p}\delta(p-\t p) -\nu \t n \rho_{\rm s} - \nu  n \t \rho_{\rm s}
	+\nu^2 n\t n \magas{C}_{00} \rt) \no
\eeqa
and
\beqa
	\lefteqn{\machain{B}_{ij} = \int \dd p\,\dd \t p\,h_i^{\rm dr} {\t h}_j^{\rm dr}\times} && \\
	&& \times \lt(
	v^{\rm eff} \rho_{\rm p} \delta(p-\t p) -
	\nu \t n v^{\rm eff} \rho_{\rm s}
	- \nu n {\t v}^{\rm eff} \t \rho_{\rm s}
	+ \nu \t n \rho_{\rm s} \averagec{q}_1
	+ \nu n \t \rho_{\rm s} \averagec{q}_1
	+ \nu^2 n\t n \magas{C}_{01}
	- \nu^2 n\t n \averagec{q}_1\magas{C}_{00}
	\rt).\no
\eeqa

{\em Proof of \eqref{chainnormal}.} We start with the expression of the conserved density given in the second equation in \eqref{avchain}. A standard result \cite{PhysRevX.6.041065,PhysRevLett.117.207201} is that, differentiating with respect to any parameter $t$ on which the occupation function $n(p)$ depends, we have
\beq
	\p_t \int \dd p\,\t h(p) n(p) h^{\rm dr}(p) = \int \dd p\,\t h^{\rm dr}(p) \p_t n(p)\, h^{\rm dr}(p).
\eeq
Therefore we find
\beq
	\p_t \averagec{q}_i = \p_t \nu \int \dd p\, n(p) h_i^{\rm dr}(p)
	+ \nu \int \dd p\, 1^{\rm dr} (p) \p_t n(p) h_i^{\rm dr}(p).
\eeq
On the other hand, differentiating the second equation in \eqref{jichain},
\beq
	\p_y \averagec{j}_i = -\p_y \averagec{q}_1 \int \dd p\, n(p) h_i^{\rm dr}(p)
	+ \int \dd p\,(v^{\rm dr}(p)-1^{\rm dr}(p)\averagec{q}_1)\p_y n(p) h_i^{\rm dr}(p).
\eeq
Since $\p_t \nu = \p_y \averagec{q}_1$, relation \eqref{veffdr} gives \eqref{chainnormal}. \eproof

\section{Derivation of classical TBA from classical scattering}\label{secttba}

We provide the derivation in the case $\nu>0$; the case $\nu<0$ would require different arguments.

Note that $\varphi(p)$ is not bounded: it diverges logarithmically at $p=0$ and $p\to\pm\infty$. The singularity at $p=0$ is however integrable, and one may assume that the states are such that large momenta are rare.

We evaluate explicitly the specific Landau potential $f$, by first evaluating the (microcanonical) large-deviation rate function $I_{\rm Landau}(\nu^{-1})$, for fixed $R$ and $N$, and performing the Legendre transform \eqref{LTlandau}. The evaluation is obtained by recasting the configuration integral over an integral on asymptotic-state data. The result is an integral over a mean-field distribution of asymptotic momenta, with two entropy contributions: one from impact parameter integrals (giving rise to a ``state density"), and the other from the combinatorics of distributing momenta given a mean-field density.

Consider the integral
\beq
	Z_{\rm micro} = \int \prod_{m=1}^N \dd x(m) \dd p(m)\,\delta(x(N)-x(1)-R)\re^{-W}
\eeq
for fixed $N$. Time evolution, from initial condition determined by the $x(m)$'s and $p(m)$'s, is a change of variable. By Liouville's theorem, the measure $\dd x(m,t)\dd p(m,t)$ is independent of $t$. The conserved quantity $W$ also is independent of $t$. Because of the condition of finite (oriented) volume $R$ at $t=0$, the particles tend, almost surely, to a scattering in-state as $t\to-\infty$. That is, almost surely, there are $p_m,x_m\in\R$ such that
\beq
	x(m,t) = p_m t + x_m + o(t),\quad m\in\{1,2,\ldots,N\} \qquad (t\to-\infty)
\eeq
with $p_m > p_{m+1}$.

The $p_m$'s are the asymptotic in-momenta, while the $x_m$'s are the impact parameters. Therefore, the expression \eqref{Was} holds. The set of variables $\{x_m,p_m:m\in\{1,\ldots,N\}\}$ is a good set parametrising the initial configuration. Indeed, the Jacobian of the transformation from the $x(m),p(m)$'s (at time $t=0$) to the $x_m,p_m$'s is in fact unity,
\beqa
	\lt|\frc{\p \{x_m,p_m\}}{\p\{x(m),p(m)\}}\rt| &=&
	\lt|\frc{\p \{x_m,p_m\}}{\p\{x(m,t),p(m,t)\}}\rt|
	\,\lt|\frc{\p \{x(m,t),p(m,t)\}}{\p\{x(m),p(m)\}}\rt| \n
	&=& \prod_m \det\lt(\frc{\p(x_m,p_m)}{\p(x(m,t),p(m,t))}\rt)\n & =& 1+o(t)
\eeqa
where in the second line we used Liouville's theorem. The asymptotic in-momenta and the impact parameters completely determine the gas's state at time 0, and form a canonical set of coordinates for it. As the Hamiltonian takes the form $H = \sum_m p_m^2/2$, the asymptotic momenta are constant, and the impact parameters evolve linearly. In integrable systems, these are ``action-angle" variables.

 It is convenient to symmetrise over all orderings of momenta, with the understanding that in the asymptotic in-state, they are positioned from left to right in decreasing order (the index $m$ in $x_m$ and $p_m$, therefore, is no longer equal to the particle's original label). Therefore, we have
\beq\label{IN}
	Z_{\rm micro} = \frc1{N!}\int \prod_{m=1}^N \dd p_m\,\re^{-\sum_{m=1}^N w(p_m)}\int\prod_{m=1}^N \dd x_m \,\delta(x(N)-x(1)-R).
\eeq
Here, $x(N)$ and $x(1)$ are complicated functions of the $x_m$'s and $p_m$'s. We are looking for
\beq\label{Rf}
	Z_{\rm micro}\asymp \re^{-RI_{\rm Landau}(\nu^{-1})}
\eeq
in the limit $N,R\to\infty$ with $R/N=\nu$ fixed.

First, we must evaluate
\beq
	S_{1,N}=S_{1,N}(p_1,p_2,\ldots) = \int\prod_{m=1}^N \dd x_m \,\delta(x(N)-x(1)-R),
\eeq
which is an entropy for the given set of asymptotic momenta configuration.

Let $R$ and $N$ be large in finite ratio. Let us consider the {\em quasiparticles}, labelled by $m$, associated to momenta $p_m$, and their trajectories, and fix all $p_m$'s. Concentrate on quasiparticle 1, and set $x_1(0)$ the point at which its trajectory crosses the zero-time slice; this is well defined up to order-1 corrections. The integrals over $x_m$ for $m\geq 2$ can be seen as providing a measure on $x_1(0)$ as a function of $x_1$. The contribution factor $S_1$ to $S_{1,N}$, due to the $x_1$ integral, is dominated by the length of the interval that $x_1$ can cover, under the condition that,  in the ``bath" of all other quasiparticles, $x_1(0)$ stays within an interval of length $R$, say $J=[0,R]$. With similar arguments for other quasiparticles, we expect
\beq
	S_{1,N} \sim \prod_{m=1}^N S_m.
\eeq
 
By factorised scattering, in order to determine $x_1(0)$ from $x_1$ we simply have to add all scattering shifts that occur as quasiparticle 1 crosses other quasiparticles in the bath to reach $x_1(0)$. At large $R$ we may assume homogeneity of the bath and that $x_1(0)$ be affected by fluctuations that are sublinear in $R$. Thus we may write $S_1 = \int \dd x_1\, \chi(x_1(0)\in[0,R])$. As $x_1$ is displaced, $x_1(0)$ is affected by a linear shift due to the linear displacement of $x_1$, plus the sum of the scattering shifts incurred as other quasiparticles are met at time 0. By homogeneity and sublinear fluctuations, this latter sum is predominantly non-fluctuating linear in $x_1$. Thus there is a slope $a>0$ such that $a x_1(0) =  x_1 + b$ and
\beq\label{S1a}
	S_1= Ra.
\eeq
At $x_1=x_1^{\rm left}=-b$ such that $x_1^{\rm left}(0)=0$, quasiparticle 1's trajectory crosses a (random) set $T_{\rm left}$ of trajectories of quasiparticles in the bath. Since all quasiparticles in the bath lie within $J$, then $T_{\rm left}$ corresponds, almost surely, to collisions on quasiparticles on the left of quasiparticle 1. A collision on the left shifts quasiparticle 1's trajectory by $\varphi(p_1-p_{m})$, bringing it farther from (closer to) the left end-point if $\varphi(p_1-p_{m})>0$ ($\varphi(p_1-p_{m})<0$). Thus
\beq\label{S1b}
	x_1^{\rm left}(0) = x_1^{\rm left} +\sum_{m\in T_{\rm left}} \varphi(p_1-p_{m}).
\eeq
 At $x_1=x_1^{\rm right}=Ra-b$ such that $x_1^{\rm right}(0)=R$, quasiparticle 1's trajectory crosses a set $T_{\rm right}$ of trajectories of quasiparticles in the bath, corresponding, almost surely, to collisions on the right of quasiparticle 1. Again, a collision on the right shifts quasiparticle 1's trajectory by $\varphi(p_1-p_{m})$, bringing it farther from (closer to) the right end-point if $\varphi(p_1-p_{m})>0$ ($\varphi(p_1-p_{m})<0$). Thus
 \beq\label{S1c}
 	x_1^{\rm right}(0) = x_1^{\rm right} -\sum_{m\in T_{\rm right}} \varphi(p_1-p_{m}).
\eeq
Further, $T_{\rm left} \cup T_{\rm right} = \{2,\ldots,N\}$ is the set of all of the bath's quasiparticles. Hence,
\beq
	R = x_1^{\rm right}(0) - x_1^{\rm left}(0) = x_1^{\rm right} - x_1^{\rm left} + \sum_{m=2}^N\varphi(p_1-p_{m}) = Ra + \sum_{m=2}^N\varphi(p_1-p_{m}).
\eeq
Therefore we find
\beq\label{x1R}
	S_1 \sim R\lt(1-\sum_{m=2}^N\frc{\varphi(p_1-p_m)}R\rt).
\eeq
This is the entropy contribution due to the integral over $x_1$. Taking into account all quasiparticles,
\beq\label{S1N}
	S_{1,N} \sim R^N\prod_{m'=1}^N \lt(1-\sum_{m=1\atop m\neq m'}^N\frc{\varphi(p_{m'}-p_m)}R\rt).
\eeq

Relations \eqref{S1a}, \eqref{S1b} and \eqref{S1c}, representing the essentially linear relation between the impact parameters and particle positions at time 0, are the relations signalling integrability in the asymptotic coordinates, and allow us to identify the asymptotic momenta and impact parameters as the ``action-angle" variables; in particular, compactness in the time-zero gas coordinates translate into a simple compactness condition in the impact parameters.

We evaluate \eqref{IN} with \eqref{S1N},
\beq
	Z_{\rm micro} = \frc{R^N}{N!}\int\prod_{m=1}^N \dd p_m\,\exp-\sum_{m=1}^N \lt[w(p_m) -
	\log\lt(1+\sum_{m'=1}^N \frc{\varphi(p_m-p_{m'})}R\rt)\rt]
\eeq
in the limit $R,N\to\infty$ with $R/N=\nu$ fixed using a mean-field approximation. This is similar to, but not exactly of the form of, the mean-field integrals studied rigorously in \cite{Ro15}. We adapt the results, as similar methods can be use, and define the empirical density
\beq
	\t\rho(p)\dd p = \frc1N \sum_m\delta(p-p_m) \dd p
\eeq
which converges in measure in the large-$N$ limit. Then, with $R^N/N! \asymp \re^{N+N\log \nu}$, the result \cite[eq. 2.43]{Ro15} gives $Z_{\rm micro} \asymp \re^{-N\t I[\t\rho]}$ where $\t\rho^*$ minimises the functional
\beq
	\t I[\t\rho] = \int \dd p\,\t\rho(p)\lt[w(p) + \log\t\rho(p) - \log\lt(1+\nu^{-1}\int \dd q\,\t\rho(q)\varphi(p-q)\rt)- 1 - \log\nu \rt]
\eeq
under the constraint $\int \dd p\,\t\rho(p) = 1$. The term $\log\t\rho(p)$ comes from counting the number of momenta distributions, given a density $\t\rho(p)$. Changing to $\rho(p) = \nu^{-1}\t\rho(p)$, we have $Z_{\rm micro} \asymp \re^{-RI[\rho_{\rm p}]}$ where $\rho_{\rm p}$ minimises
\beq
	I[\rho] = \int \dd p\,\rho(p)\lt[w(p) + \log\rho(p) - 1 - \log\lt(1+\int \dd q\,\rho(q)\varphi(p-q)\rt)\rt]
\eeq
under the constraint
\beq\label{constr}
	\int \dd p\,\rho(p) = \nu^{-1}.
\eeq
Therefore,
\beq
	I_{\rm Landau}(\nu^{-1}) = I[\rho_{\rm p}].
\eeq

Introducing the Lagrange parameter $\lambda$ for the constraint \eqref{constr}, the minimisation gives $\delta I[\rho]/\delta \rho(p)|_{\rho_{\rm p}} = \lambda$, which is
\beq
	\lambda = w(p) + \log\rho_{\rm p}(p) -\log\lt(1+\int \dd q\,\rho(q)\varphi(p-q)\rt)
	- \int \dd q\,\frc{\rho_{\rm p}(q)\varphi(p-q)}{1+\int \dd r\,\rho(r)\varphi(q-r)}.
\eeq
Denoting
\beq\label{prdef}
	\re^{-\ep(p)} = \frc{\rho_{\rm p}(p)}{1+\int \dd q\,\rho(q)\varphi(p-q)}
\eeq
this simplifies to
\beq\label{prep}
	\ep(p) = w(p) - \lambda - \int \dd q\,e^{-\ep(q)}\varphi(p-q).
\eeq

Finally, the Legendre transformation \eqref{LTlandau} gives
\beq
	\mu = \frc{\dd I_{\rm Landau}(\nu^{-1})}{\dd \nu^{-1}} =
	\int \dd p\,\frc{\delta I[\rho]}{\delta \rho(p)}\Big|_{\rho=\rho_{\rm p}} \frc{\dd \rho_{\rm p}(p)}{\dd \nu^{-1}} = 
	\lambda \int \dd p\,\frc{\dd \rho_{\rm p}(p)}{\dd \nu^{-1}} = \lambda \frc{\dd}{\dd \nu^{-1}}
	\int \dd p\,\rho_{\rm p}(p) = \lambda
\eeq
as well as
\beq\label{fmu}
	\frc{\p f}{\p \mu} = -\nu^{-1} = - \int \dd p\,\rho_{\rm p}(p)
\eeq
and
\beq\label{frho}
	\frc{\delta f}{\delta w(p)} = \frc{\delta I[\rho]}{\delta w(p)}\Big|_{\rho=\rho_{\rm p}}
	= \rho_{\rm p}(p).
\eeq
Equations \eqref{fmu} and \eqref{frho} are expressions \eqref{avgas}, definition \eqref{prdef} is \eqref{rhoprhos}, and \eqref{prep} is \eqref{ep}. Together with \eqref{qgas}, this shows \eqref{f}. Note also that the quantity $\rho_{\rm s}$ defined in \eqref{statedensity} arises from the impact-parameter entropy, Eq. \eqref{x1R}.

\section{Conclusion}\label{sectconclu}


In this paper, we have constructed the GGEs and Euler GHD of the Toda system, both in the viewpoint of the gas and of the chain. We have obtained explicit expressions for the average densities and currents, the Euler equation, the explicit normal modes of the hydrodynamics as well as some hydrodynamic matrices both in the gas and the chain. From this, using the results in the literature, more can be evaluated exactly, for instance the exact profiles in the partitioning protocol \cite{PhysRevX.6.041065,PhysRevLett.117.207201}.

The approach was based on a scattering analysis of the Toda gas. The results therefore take the standard GHD form for the gas hydrodynamics, based on the classical TBA framework, and more results in the GHD literature are, conjecturally, immediately applicable, such as the exact Drude weights \cite{SciPostPhys.3.6.039}, the Euler-scale correlation functions \cite{doyoncorrelations}, the large-deviation theory for transport \cite{Myers-Bhaseen-Harris-Doyon}, and the exact hydrodynamic force terms \cite{DY} and hydrodynamic diffusive corrections \cite{dNBD}.

Results in the chain are obtained by using the equivalence of ensembles; they are expected to hold at the Euler scale, and take a slightly different form. As a consequence, some of the results in the literature have to be re-worked out, and perhaps most crucially the diffusion operator needs to be studied. One way forward is to do the scattering analysis for the chain. In this case, the asysmptotic states are the solitons and radiative corrections. Soliton and radiation shifts are known. In fact, the classical TBA describing baths of solitons and radiative modes was obtained by taking a particular semi-classical limit of the quantum Toda TBA, separating large- and small-quasimomentum modes  \cite{TI86}. This was done for thermal states, but the extension to GGEs is possible, giving the specific Gibbs free energy for the ensemble ${\cal E}_{\rm Gibbs}$. From this, one may in principle use the machinery of GHD and obtain the chain hydrodynamics directly, including diffusive corrections. It would be interesting to do this and to analyse how this approach reproduces the Euler scale results we have obtained.

The approach we have used also emphasises the freedom in choosing scattering states at the basis of TBA. The set of asymptotic excitations and their scattering, whence the TBA, depends on the choice of vacuum -- here that of the gas, without particles, instead of that of the chain, with a uniform density of particles. This is a general concept which may be useful to exploit.

The classical gas TBA can also be obtained by a semi-classical analysis, done in \cite{T84,O85}. We have given an independent, explicit derivation of the classical TBA from classical scattering and the definition of the thermodynamic ensemble. This therefore explains purely from classical scattering the semiclassical result, in particular identifying the quasiparticles of classical TBA as the velocity tracers of the Toda particles themselves. The analysis of the semi-classical limit provided in \cite{BCM19} (which appeared shortly after the first version of the present paper) discusses further its relation to the quasiparticles.

The thermodynamics of the Toda chain was also obtained in \cite{SpoToda} by making a connection with the Dumitriu-Edelman random matrix theory model \cite{DE02}. The latter model was shown \cite{DE02} to produce eigenvalue distribution in the $\beta$-ensemble, and in \cite{SpoToda} the mean-field regime $\beta = O(1/N)$ is used. This relation opens many research directions. In particular, the derivation of the classical TBA we have provided effectively gives rise to a new derivation of the $\beta$-ensemble statistics of the Dumitriu-Edelman model in a certain limit, hinting at a scattering underpinning. The  $\beta$-ensemble form is valid in the Dumitriu-Edelman model also before the thermodynamic limit is taken; is there a modification of the Toda system that would give a $\beta$-ensemble distribution of asymptotic momenta in finite volumes?

Finally, we point out two aspects which may be useful for applications to other models. First, the derivation we have given in subsection \ref{ssectscat} for factorised scattering, and for the concept of quasiparticle, based on an adaptation to classical particles of an argument by Parke \cite{Parke80}, necessitates {\em a single conserved quantity} associated to a higher power of momentum (up to appropriate continuity arguments). This emphasises the fact that a single higher conserved charge is often sufficient for the most crucial aspects of integrability. Second, in one dimension, two-body scattering processes preserve the set of momenta. Therefore, by the arguments in subsection \ref{ssectscat}, the factorised scattering structure holds as well in {\em low-density gases of non-integrable models}. It is clear also that the full derivation of the classical TBA presented in section \ref{secttba} holds without modification. Therefore one expects GHD to describe such gases, at least at short times, before higher-body processes start being important. It would be very interesting to study further such ``pre-thermalisation" effects in low-density non-integrable gases.

\medskip
\medskip
{\bf Acknowledgments.} I am most grateful to Herbert Spohn for suggesting this problem to me and for the many insightful discussions. I also thank Yan Fyodorov, and the Disordered System and Theoretical Physics Groups at King's College London, for comments and interest. H. Spohn reported to me the observation, made by Xiangyu Cao and Vir Bulchandani, that $f$ has two branches. This work was funded by the Royal Society under a Leverhulme Trust Senior Research Fellowship, ``Emergent hydrodynamics in integrable systems: non-equilibrium theory", ref. SRF\textbackslash R1\textbackslash 180103. I thank the Centre for Non-Equilibrium Science (CNES) and the Thomas Young Centre (TYC).

\vspace{2cm}

\noindent{\Large\bf Appendix}

\appendix

\section{Standard forms for chain hydrodynamics}\label{app}

Although the expressions \eqref{avchain}, \eqref{j0chain} and \eqref{jichain} are not immediately in the standard form prescribed by TBA and GHD, it is possible to bring them to this form. The standard form depends on the parametrisation of the quasiparticle chosen. The above expressions in fact amount to redefining the basic quantities so as to see $p$ not as the momentum, but as a parametrisation of the ``actual" chain momentum. Both the chain momentum and energy take a form, as functions of $p$, which account for the change of coordinate from the gas to the chain, and for the change of (local) Galilean frame.

First, the chain momentum function, parametrised by $p$, is given by
\beq
	p^{\rm c}(p) = \nu (p-\averagec{q}_1).
\eeq
The $p$-independent shift does not affect any of the equations below, however it is natural to keep it within this definition. This parametrisation is in fact explicitly state dependent, in particular $\nu$ depends on the state itself. This however does not affect the formal structure in the TBA technology. It is also natural to define
\beq
	\rho_{\rm p}^{\rm c}(p) = \nu \rho_{\rm p}(p)
\eeq
as the quasiparticle density per unit phase-space with respect to lengths on the chain, instead of oriented lengths in the gas; this is a positive quantity. The occupation function and jump function do not change, $n^{\rm c}(p) = n(p)$ and $\varphi^{\rm c}(p,p') =\varphi(p-p')= 2\log|p-p'|$. As a consequence, the dressing operation also is left unchanged, $h^{\rm dr}_{\rm c}(p) = h^{\rm dr}(p)$. The standard TBA relations hold, with the ``state density"
\beq
	\rho_{\rm s}^{\rm c}(p) = \lt(\frc{\dd p^{\rm c}}{\dd p}\rt)^{\rm dr}(p) =
	\frc{\dd p^{\rm c}(p)}{\dd p} + \int \dd p'\, \varphi(p,p')\,\rho_{\rm p}^{\rm c}(p')
\eeq
and $\rho_{\rm p}^{\rm c}(p) = n(p) \rho_{\rm s}^{\rm c}(p)$. One can check that the first relations in \eqref{avchain} are equivalent to
\beq\label{avchaintba}
	\averagec{q}_i = \int \dd p\,h_i(p) \rho_{\rm p}^{\rm c}(p)
	= \int \dd p^{\rm c}(p)\,h_i^{\rm dr}(p) n(p)\quad (i\in\N),
\eeq
which are the standard TBA equations for average densities, in the chain parametrisation.

Second, we define the chain energy function as
\beq
	E^{\rm c}(p) = \frc12(p - \averagec{q}_1)^2.
\eeq
That is, the energy of a quasiparticle $p$, from the chain point of view, is that of a particle in the rest frame, where the average momentum vanishes. Equivalently, it is that of a particle of momentum $p+\averagec{j}_{\b 0}$, which takes into account the contribution to the momentum due to the soliton current. For the energy, one does not multiply by the factor $\nu$, as the energy is not affected by the stretch in space. The chain effective velocity $v^{\rm eff, c}(p)$ solves, according to the standard GHD equation,
\beq\label{veffch}
	\frc{\dd p^{\rm c}(p)}{\dd p}v^{\rm eff,c}(p) = \frc{\dd E^{\rm c}(p)}{\dd p} + \int \dd p'\,\varphi(p,p')\, \rho_{\rm p}^{\rm c}(p')(v^{\rm eff, c}(p')-v^{\rm eff,c}(p))
\eeq
and can be written equivalently as
\beq\label{veffdressingch}
	 v^{\rm eff,c}(p) = \frc{(\dd E^{\rm c}/\dd p)^{\rm dr}(p)}{(\dd p^{\rm c}/\dd p)^{\rm dr}(p)}.
\eeq
It is simple to check that this gives \eqref{veffchain}. One can then verify that \eqref{jichain} are equivalent to
\beq
	\averagec{j}_i = \int \dd p\,h_i(p)v^{\rm eff,\rm c}(p)\rho_{\rm p}^{\rm c}(p) 
	= \int \dd E^{\rm c}(p)\,n(p) h_i^{\rm dr}(p),
\eeq
which are the standard TBA equations for average currents, in the chain parametrisation.

\end{document}